\documentclass[iop,apj]{emulateapj}
\journalinfo{The Astrophysical Journal} 
\usepackage{epstopdf}
\usepackage{amsmath,graphicx}
\usepackage{tensor}
\usepackage{mathrsfs}
\usepackage{color}
\usepackage{textcomp}
\usepackage{gensymb}
\usepackage{hyperref}
\usepackage{multirow}
\shorttitle{Cosmic Ray Diffusion coefficients from 3-D SW simulation}
\shortauthors{Chhiber et al.}

\begin{document}

\title{Cosmic Ray Diffusion coefficients throughout the inner heliosphere from global solar wind simulation}
\author{R.~Chhiber$^1$, P.~Subedi$^1$, A.V.~Usmanov$^{1,2}$, W.H.~Matthaeus$^1$, D. Ruffolo$^3$, M.L.~Goldstein$^2$, and T.N.~Parashar$^1$}
\affil{
$^1$Bartol Research Institute and Department of Physics and Astronomy, University of Delaware, Newark, DE 19716, USA \\
$^2$ NASA Goddard Space Flight Center, Greenbelt, MD 20771, USA \\
$^3$ Department of Physics, Faculty of Science, Mahidol University, Bangkok 10400, Thailand}

\begin{abstract}
We use a three-dimensional magnetohydrodynamic simulation of the solar wind to calculate cosmic ray diffusion coefficients throughout the inner heliosphere ($2~R_\odot - 3$ AU). The simulation resolves large-scale solar wind flow, which is coupled to small-scale fluctuations through a turbulence model. Simulation results specify background solar wind fields and turbulence parameters, which are used to compute diffusion coefficients and study their behavior in the inner heliosphere. The parallel mean free path is evaluated using quasi-linear theory, while the perpendicular mean free path is determined by non-linear guiding center theory with the random ballistic interpretation. Several runs examine varying turbulent energy and different solar source dipole tilts. We find that for most of the inner heliosphere, the radial mean free path (mfp) is dominated by diffusion parallel to the mean magnetic field; the parallel mfp remains at least an order of magnitude larger than the perpendicular mfp, except in the heliospheric current sheet, where the perpendicular mfp may be a few times larger than the parallel mfp; in the ecliptic region, the perpendicular mfp may influence the radial mfp at heliocentric distances larger than 1.5 AU; our estimations of the parallel mfp in the ecliptic region at 1 AU agree well with the Palmer ``consensus" range of $0.08 - 0.3$ AU; solar activity increases perpendicular diffusion and reduces parallel diffusion; the parallel mfp mostly varies with rigidity $(P)$ as $P^{.33}$, and the perpendicular mfp is weakly dependent on $P$; the mfps are weakly influenced by the choice of long wavelength power spectra.
\end{abstract}


\section{Introduction}

The interaction of energetic particles with the solar wind is a topic of wide interest in space physics and astrophysics. Several varieties of charged particles populate the heliosphere, including energetic particles originating at the sun (solar energetic particles, or SEPs) and galactic cosmic rays (GCRs) that enter the heliosphere uniformly and nearly isotropically from the outside \citep{Kunow1991book}. These cosmic rays (CRs) are strongly guided and scattered by the solar wind and the turbulent fluctuations that transport with it \citep{parker1956modulation,parker1964scattering,jokipii1966cosmic}. As such, the study of the origin and transport of cosmic rays is an important problem in heliospheric physics, with implications ranging from space weather and exploration to fundamental space plasma physics  \citep{jokipii1971review,fisk1978interactions,Kunow1991book}. The effects of these energetic particles on the health of astronauts \citep{Parker2005SWE} and the well-being of electronic components in spacecraft \citep{Tylka1997} are an immediate concern. In addition, the accuracy with which we can understand CR propagation also provides a testbed for energetic particle transport in numerous space and astrophysical applications \citep{Kulsrud1969ApJ,droge2003}. The solar wind provides us with an opportunity to observe, at close range, the behavior of energetic particles in random, turbulent magnetic fields \citep{bruno2013LRSP}. Such fields are ubiquitous in astrophysical systems \citep{Candia2004}, and the insights we glean from studies of CRs in the heliosphere can potentially find application elsewhere in the universe. Finally, observations of cosmic rays can also serve as probes into solar activity and solar wind structure, as CR variations are seen to be correlated with solar and geomagnetic activity \citep{snyder1963}.
  
Theories of the modulation of cosmic rays in the heliosphere attempt to explain the observed temporal and spatial variation in their spectra \citep{fisk1978interactions,potgieter2013solar}, and for that purpose, require a knowledge of the cosmic ray diffusion tensor. In fact, one of the key challenges in solving the Parker CR transport equation \citep{parker1965PSS} is the inadequate knowledge of the spatial, temporal, and rigidity dependence of the components of the diffusion tensor. In turn, the specification of this tensor through the heliosphere requires an understanding of two topics. First, a theoretical understanding of the diffusion process itself is needed, which would lead to predictions of the structure of the diffusion tensor itself. Equally important is the knowledge of the large scale flows and electromagnetic field in the plasma, and the distribution of background solar wind turbulence in which the particles are scattered. The present approach permits three dimensional, and (in principle) time-varying calculation of all three of these properties (diffusion  tensor, large scale flow, large scale electromagnetic field) to be computed in a single model.

The formal structure of the diffusion tensor involves diagonal components corresponding to diffusion parallel and perpendicular to the interplanetary magnetic field (IMF), as well as off-diagonal components describing perpendicular drifts (e.g., \citealt{moraal1976SSRv,minnie2007ApJ}). While quasi-linear theory \citep{jokipii1966cosmic} extended to include time-dependent and non-linear corrections \citep{goldstein1976ApJ,bieber1994proton,droge2003} provides a relatively good accounting of parallel diffusion, theories of perpendicular diffusion have faced the challenge of accounting for non-linear effects such as transfer of particles across field lines, backscatter from parallel diffusion, and field-line random walk \citep{jokipii1966cosmic,Giacalone1999ApJ}. The non-linear guiding center (NLGC) theory (\citealt{matthaeus2003nonlinear}; see also \citealt{shalchi2009}) accounts for the above, and is further improved by the random ballistic interpretation of \cite{ruffolo2012random}. In  the current work we focus on the parallel and perpendicular and diffusion coefficients; the drift motion could be a topic for future work. 

Since turbulent fluctuations are responsible for scattering CRs, the diffusion theories mentioned above typically involve turbulence parameters such as the energy of the random magnetic fluctuations and correlation scales. In the solar wind, low-frequency turbulence evolves via a non-linear cascade, while also being transported and processed by the large-scale radially expanding solar wind. At all but the smallest scales, these processes are well described by magnetohydrodynamic (MHD) models \citep{marsch1989dynamics,zhou1990transport}. Over the years, turbulence models have incorporated simplifying assumptions relevant to the solar wind, yielding increased tractability of the governing equations \citep{zank1996evolution,matthaeus1999turbulence}. The increased sophistication of the models and improvements in computational power have led to numerical simulations yielding good agreement with \textit{Voyager, Ulysses, Helios}, and \textit{WIND} observations \citep{breech2008turbulence,usmanov2011solar}. These turbulence models have also been used to study the propagation of coronal mass ejections \citep{Wiengarten2015ApJ}. Extensions to the \cite{breech2008turbulence} model have been developed \citep{Oughton2011JGRA116,Zank2012ApJ745,Zank2017ApJ835}, and applied to the inhomogeneous solar wind \citep{Shiota2017ApJ837}.

Our strategy for evaluating the CR diffusion coefficients through the inner heliosphere consists of two steps: first, specification of the relevant turbulence parameters based on a global solar wind model, and second, evaluation of the CR diffusion coefficients using the specified heliographic distribution of turbulence. For the first step, we deduce turbulence parameters from a global, three-dimensional (3-D) MHD simulation of the solar wind \citep{usmanov2014three}. 

The spatial resolution that can be realistically achieved in such simulations cannot resolve the small-scale fluctuations that cause scattering of CRs. For instance, the spatial resolution of our simulation, at 1 AU, can be estimated as roughly 0.03 AU. However, the mean free path, at 1 AU, for scattering perpendicular to the mean magnetic field has been estimated to be as low as 0.001 AU \citep{zhang2009ApJ,pei2010cosmic}, and the correlation scale of the turbulence has been estimated to be around 0.007 AU \citep{matthaeus2005,bruno2013LRSP}. This is where our turbulence model for the ``sub-resolution" physics comes in. Our simulation explicitly resolves the large-scale, mean solar wind bulk flow, which is coupled to small-scale inhomogeneities by means of an MHD-Reynolds-averaged Navier-Stokes \citep[RANS; see, e.g.,][]{mccomb1991physics} model for the random fluctuations. The simulation has been well-tested, and gives reasonable agreement with many spacecraft observations of large-scale solar wind fields, turbulence parameters (energy, cross helicity and  correlation scale), as well as the temperature, for varying heliocentric distance, and where feasible, varying helio-latititude \citep{usmanov2011solar,usmanov2012three,usmanov2016four}. In recent ``applied" work, the simulation has been used to study the collisional age of the solar wind plasma \citep{chhiber2016solar}, and we view the present work as a continuation of such application-oriented studies.

Once the turbulence parameters are specified through the model heliosphere, for the second step of our calculation, we use, as a starting point, fairly standard, well-tested formalisms for parallel and perpendicular diffusion coefficients - quasi-linear theory \citep{jokipii1966cosmic,bieber1995diffusion,zank1998radial} to compute the parallel component of the diffusion tensor, and the random ballistic decorrelation (RBD) interpretation of NLGC theory \citep{matthaeus2003nonlinear,ruffolo2012random} for perpendicular diffusion.     
    
Previous studies of the heliographic dependence of the CR diffusion coefficients include work based on both WKB models for Alfv\'{e}n waves \citep{volk1974spatial,morfill1979latitude}, and models for strong turbulence \citep{bieber1995diffusion,zank1998radial,pei2010cosmic}. The present work builds on these studies, but also makes some significant departures, motivated and enabled by recent advances in diffusion theory and sophistication of solar wind simulations. The major points of departure from previous work are listed below:

1. We use a fully 3-D global simulation of the solar wind that provides us with a reliable and self-consistent model heliosphere. Previous work has used one-dimensional (1-D) radial evolution models with spherical symmetry, with shear-driving effects included through a model \citep{zank1998radial,pei2010cosmic}. Thus, while examining latitudinal dependence of the diffusion tensor, these studies implicitly assume that they are far from regions with significant latitudinal gradients. In contrast, three dimensionality improves the physical authenticity of the simulation by explicitly including shear-driving effects on the flow across latitudes, and leads to improved data-visualization through two-dimensional (2-D) contour plots. A similar 3-D approach has been recently used in \cite{guo2016ApJ} to study the propagation of GCRs from 0.3 AU to the termination shock.

2. The computation of the CR diffusion tensor requires specification of the background solar wind speed, and the underlying large-scale heliospheric magnetic field. Previous work \citep{bieber1995diffusion,zank1998radial,pei2010cosmic} used a radially constant solar wind speed with some latitudinal variation, and a Parker-spiral type magnetic field model. However, the use of a prescribed model for the background fields has been found inadequate \citep{Reinecke1997AdSpR,steenberg1997alternative}, and instead we use the large-scale, resolved flow from our MHD-RANS simulation. This provides a complete specification of the background large-scale fields, with spatial variation that has been found to agree well with observations \citep{usmanov2014three}.

3. We examine the diffusion coefficients at radial distances between 2 $R_\odot$ and 3 AU, where $R_\odot$ denotes a solar radius. We are not aware of any other similar study that has probed regions this close to the sun, which are of prime interest for SEP propagation, space weather, and for upcoming spacecraft missions, including Solar Probe Plus. Resolving this entire domain ($2~R_\odot - 3$ AU) in one simulation is a challenge, as modeling approximations that are appropriate very close to the sun may not be valid at larger heliocentric distances. Furthermore, the timescales associated with the different domains are disparate \citep{hundhausen1972coronal,Tu1995SSRv,bruno2013LRSP}. We use an approach where the computational domain is split into three regions: inner ($1-20~R_\odot$), intermediate ($20 - 45~R_\odot$), and outer ($45~R_\odot - 3$ AU). The inner and intermediate regions employ a WKB Alfv\'{e}n wave model, and the outer region solves a full turbulence transport model, with the inner boundary conditions for each region being provided by the preceding one \citep{usmanov2014three}. 

4. A magnetic dipole with its tilt (relative to the solar rotation axis) varying through the solar activity cycle is a first and rough approximation for the solar magnetic field \citep{babcock1961ApJ133}. We examine the effect of changing the tilt of the source solar dipole by using simulations with a dipole untilted with respect to the solar rotation axis, and a dipole with {30\degree} tilt, in contrast to previous work employing axisymmetric solar wind parameters \citep{zank1998radial,pei2010cosmic}. The tilt of the solar dipole and the warping of the helispheric current sheet \citep{smith2001JGRsheet} indicate high levels of solar activity \citep{heber2006}, which is a factor of interest since CR intensity is anticorrelated to solar activity levels \citep{forbush1954JGR,fisk1978interactions}. We note here that previous work that examined the effect of solar activity on CR-intensity variation \citep{jokipii1995conf} did not include turbulence modeling, and here we examine how varying turbulence levels influence the diffusion coefficients. 

5. The perpendicular diffusion coefficient has been previously evaluated using the ``BAM" model \citep{bieber1997perpendicular} by \cite{zank1998radial}, and the NLGC theory \citep{matthaeus2003nonlinear} by \cite{pei2010cosmic} and \cite{zank2004JGR}. Recently, the NLGC theory has been reinterpreted by \cite{ruffolo2012random}, and their RBD theory yields a significantly improved agreement with numerical experiments, for magnetic fluctuation amplitudes comparable to the large-scale magnetic field. This makes it very well suited for application to the solar wind, where the IMF includes a strong fluctuating component \citep{belcher1969JGR,marsch1991pihp}, and we use the RBD theory to derive a new expression for the perpendicular diffusion coefficient. 

6. With the above improvements, the present approach departs significantly from both SEP studies \citep[e.g.,][]{zhang2009ApJ} and GCR modulation studies \citep[e.g.,][]{Engelbrecht2013} that have used relatively simplified assumptions in one or more of the above categories, such as semiempirical diffusion coefficients and simple scalings with magnetic field magnitude.

The outline of the paper is as follows: We describe the form of the CR diffusion tensor in Section 2, and briefly discuss the turbulence model and the simulation in Section 3. Section 4 presents the heliographic distribution of the diffusion coefficients. In an Appendix we briefly describe how other types of diffusion coefficients might be estimated using similar approaches.

\section{Cosmic Ray Diffusion Tensor}

The CR diffusion tensor, $\kappa_{ij}$, describes the scattering of CRs by random fluctuations in the IMF. It may be expressed as \citep{parker1965PSS,jokipii1970ApJ}
\begin{equation}\label{eq:kappa}
\kappa_{ij} = \kappa_\perp \delta_{ij} + \frac{B_i B_j}{B^2} (\kappa_\parallel - \kappa_\perp) + \epsilon_{ijk} \kappa_A \frac{B_k}{B},
\end{equation}
where $\mathbf{B}$ is the mean IMF, $\delta_{ij}$ is the Kronecker delta, and $\epsilon_{ijk}$ is the Levi-Civita symbol. This work presents calculations of $\kappa_\parallel$ and $\kappa_\perp$, which are the diagonal components of the diffusion tensor parallel and perpendicular, respectively, to the mean IMF. 

The present work does not calculate $\kappa_A$, which can describe particle drifts under the influence of large-scale gradients and curvature in the IMF.  Our results are directly relevant to the outward propagation of SEPs, for which $\kappa_\parallel$ and $\kappa_\perp$ are needed to describe how the SEP distribution spreads in the parallel and perpendicular directions, whereas over the short time scale of the SEP outflow the drifts may mainly shift the lateral distribution over a small angle. The lateral distribution of particle injection is often unknown, and the effects of drifts are often neglected, though \cite{marsh2013ApJ} argue that they should be considered.  Both diffusion and drifts are considered to be important to the modulation of GCR with the solar cycle and the small gradients in GCR density \citep{moraal1976SSRv,jokipii1981ApJ}, though these processes take place over a wider region than considered in the present work ($r\le3$ AU).

We shall also examine the radial diffusion coefficient
\begin{equation}\label{kappa_r}
\kappa_{rr} \equiv \kappa_\parallel \cos^2 \Psi  + \kappa_\perp \sin^2 \Psi,
\end{equation}
which is of particular relevance to models of solar modulation of CRs. Here, $\Psi$ is the ``winding" angle between the IMF and the radial direction. Following previous work, we define mean free paths, $\lambda_{\parallel,\perp}$, that are equivalent to the diffusion tensor through 
\begin{equation}
\lambda_{\parallel,\perp} \equiv 3 \kappa_{\parallel,\perp}/v,
\end{equation}
where $v$ is the particle speed.

We note that in the present work we use the large-scale flow from our simulation to specify $B$ and $\Psi$ as spatially varying fields through the 3-D heliosphere. This is in contrast to previous studies \citep{bieber1995diffusion,zank1998radial,pei2010cosmic}, where $B$ and $\Psi$ were specified through a Parker-type model and a radially constant solar wind speed (to compute $\Psi$). However, the features of the IMF have a major influence on CR transport, and a Parker-type field is an oversimplification, particularly at high heliolatitudes (See \cite{heber2006} for an overview of suggested modifications to the Parker field). Moreover, the use of a-priori prescribed background fields in modulation studies has been held responsible for restricting the diffusion tensor to values that preclude agreement of models with observations \citep{Reinecke1997AdSpR,steenberg1997alternative}, and the present work makes a significant improvement in this regard.   

\subsection{Parallel mean free path}

In determining the parallel mean free path (mfp), the turbulence ``geometry", i.e., the distribution of energy over parallel and perpendicular wavevectors, is a controlling factor. Observations \citep{bieber1994proton} show that a pure ``slab" model of heliospheric turbulence \citep{jokipii1966cosmic} underestimates the parallel mfp. In the slab model, the magnetic fluctuations are polarized perpendicular to the mean field and their wave-vectors are parallel to the mean field. \cite{bieber1994proton} find that a composite model with a dominant 2-D part (fluctuations and their wave-vectors both perpendicular to the mean field) and a minor slab part provides a better approximate parametrization of the turbulence and an improved description of the observed mean free paths. Furthermore, theoretical studies and observations \citep{matthaeus1990JGR,zank1992waves,zank1993nearly,bieber1996dominant,ghosh1997anisotropy} suggest that around 80\% of magnetic fluctuation energy in the inertial range should reside in the 2-D component, with the rest in the slab component. 

In the following, we take the $z$-component along the mean field. Considering parallel diffusion first, we note that in quasilinear theory the 2-D fluctuations are effectively invisible to CRs resonating with the turbulence, and the scattering by slab fluctuations (assumed to be axisymmetric) is described by the parallel mfp \citep{zank1998radial} 

\begin{equation}\label{eq:mfp_p}
\begin{aligned}
\lambda_\parallel &= 6.2742 \frac{B^{5/3}}{\langle b^2_s \rangle} \left(\frac{P} 
                  {c}\right)^{1/3}  \lambda^{2/3}_{s}          \\
                  &\times \left[1 + \frac{7A/9}{(1/3 + q)(q + 7/3)}\right],
\end{aligned}
\end{equation}

where 
\begin{equation}
A = (1+s^2)^{5/6} - 1,
\end{equation}
\begin{equation}
q = \frac{5s^2/3}{1+s^2-(1+s^2)^{1/6}},
\end{equation}
\begin{equation}
s = 0.746834 \frac{R_L}{\lambda_s},
\end{equation}
and a model 1-D Kolmogorov spectrum is assumed, with a power spectrum of the form $\tilde{P}(k_\parallel) \propto (1+k_\parallel \lambda_s)^{-5/6}$. Here $c$ is the speed of light, $R_L = P/Bc$ the particle Larmor radius, $\langle b_s^2 \rangle$ the variance of the slab geometry fluctuation, $P \equiv \tilde{p}c/Ze$ the particle rigidity ($\tilde{p}$ and $Ze$ are the particle momentum and charge, respectively), $k_\parallel$ is the wave vector parallel to the mean field, and $\lambda_s$ the correlation length for slab turbulence. Equation \eqref{eq:mfp_p} is valid at rigidities ranging from from 10 MV to 10 GV \citep{zank1998radial}. At larger heliocentric distances, the fractional term in braces becomes significant due to high rigidity particles resonating with fluctuations in the energy containing range instead of the inertial range. This is discussed further below in the context of rigidity dependence of the mfps (Section 4.4).

\subsection{Perpendicular mean free path}
Perpendicular diffusion is often not considered as important as parallel diffusion in energetic particle studies, because it is usually inferred that $\lambda_\perp << \lambda_\parallel$ \citep{palmer1982RvGSP}. However, \cite{dwyer1997ApJ} found that for strong particle enhancements related to corotating interaction regions, $\lambda_\perp/\lambda_\parallel$ rose to $\sim1$ in the fast solar wind stream arriving after the stream interface.  Using data from the \textit{Ulysses} spacecraft during the SEP event of 2000 Jul 14, \cite{zhang2003ApJ} inferred $\lambda_\perp/\lambda_\parallel\approx 0.25$. Our 3-D model inner heliosphere provides an opportunity to examine the domains where perpendicular diffusion can be comparable with parallel diffusion.

Quasi-linear theory \citep{jokipii1966cosmic} provides a physically appealing description of perpendicular diffusion in terms of the diffusive spread of magnetic field lines, with the gyrocenters of charged particles following the field lines. Other approaches have considered the relationship between $\kappa_\perp$ and $\kappa_\parallel$ \citep{Axford1965P&SS,Gleeson1969P&SS}, and applied the Taylor-Green-Kubo formulation (BAM, \citealt{bieber1997perpendicular}) to the problem. However, the field line random walk (FLRW) approach \citep{jokipii1966cosmic} overestimates the strength of the diffusion, while BAM predicts diffusion that is weaker than that observed in numerical experiments \citep{Giacalone1999ApJ,mace2000ApJ}. The NLGC theory \citep{matthaeus2003nonlinear} accounts for both the random walk of the field lines and the influence of parallel scattering, and shows good agreement with both observations \citep{bieber2004GRL} and simulations, with the NLGC results bracketed by the FLRW and BAM results \citep{matthaeus2003nonlinear}.  

Recent work \citep{ruffolo2012random} has reinterpreted NLGC by replacing the diffusion of gyrocenter trajectories with a random ballistic decorrelation (RBD), where the guiding center motion is approximated as ballistic (i.e., with constant velocity) between scattering events. The RBD-modified theory agrees with numerical simulations over a wider range of fluctuation amplitudes than the original NLGC, specifically for fluctuations comparable in size to the large-scale field. This makes it particularly suited for application to the solar wind \citep{belcher1969JGR,marsch1991pihp}. Other improvements to NLGC have also been developed \citep[see, e.g.,][]{shalchi2009}.

The phenomenon of ``backtracking" due to parallel scattering causes a particle to reverse its motion along the field line, thus retracing its steps over a certain time-span. This leads to a negative $v_x$-correlation ($v_x$ being a component of the particle's velocity perpendicular to the mean field), which results in a reduction in the running perpendicular diffusion coefficient. With this backtracking correction, RBD yields the following perpendicular diffusion coefficient \citep{ruffolo2012random}:
\begin{equation} \label{eq:rbd}
\kappa_{\perp} = \frac{a^2 v^2}{6 B^2} \sqrt{\frac{\pi}{2}} 
\int_0^\infty 
\frac{S_{2} (k_\perp) \text{Erfc}(\alpha) 2 \pi k_{\perp} dk_{\perp}}
{k_\perp \sqrt{\langle \tilde{v}_x^2 \rangle}},
\end{equation}
where $a^2 = 1/3$, $v$ is the particle speed, $\tilde{v}_x$ is the $x$-component of the guiding center velocity, $S_{2}$ is the 2-D axisymmetric turbulent fluctuation spectrum, Erfc is the complementary error function, and $k_\perp$ is the component of the wave-vector perpendicular to the mean magnetic field. We also have
\begin{equation} \label{eq:alpha}
\alpha = \frac{v^2}{3 \kappa_\parallel k_\perp \sqrt{2 \langle \tilde{v}_x^2 \rangle}},
\end{equation}
and
\begin{equation}
\langle \tilde{v}^2_x \rangle = \frac{a^2 v^2 b^2} {6 B^2},
\end{equation}
where $b^2$ is the combined variance of the 2-D and slab magnetic fluctuations: $b^2 = \langle {b}_2^2 \rangle+\langle {b}_s^2 \rangle $. Note that in Equation \eqref{eq:rbd}, the slab turbulence spectrum does not appear. This is because we follow the suggestion by \cite{Shalchi2006A&A} that the direct contribution of the slab component to perpendicular transport is subdiffusive, and therefore the slab term should not contribute to Equation \eqref{eq:rbd}. This hypothesis has been supported by simulations \citep{ruffolo2012random,2012AGUFMSH21A2188R}, and accordingly, has been adopted in the present work as well. Slab fluctuations can, however, still influence $\kappa_\perp$ through $\kappa_\parallel$, which appears in Equation \eqref{eq:alpha} for $\alpha$, and $\langle \tilde{v}^2_x \rangle$.

The 2-D power spectrum may be expressed as a power law \citep{matthaeus2007spectral}
\begin{equation} \label{eq:twod1}
S_{2} (k_\perp \leq 1/\lambda_{2}) = C_{2} \langle b_{2}^2 \rangle \lambda_{2}^2
                                       (\lambda_{2} k_\perp)^p,
\end{equation}
\begin{equation} \label{eq:twod2}
S_{2} (k_\perp > 1/\lambda_{2}) = C_{2} \langle b_{2}^2 \rangle \lambda_{2}^2
                                       (\lambda_{2} k_\perp)^{-\nu -1},
\end{equation}
where $\lambda_2$ is the 2-D correlation scale, $C_2$ is a normalization constant,  $\langle b_{2}^2 \rangle$ is the variance of the 2-D turbulent fluctuations, and $p$ is a power index that takes on integral values that correspond to different power spectra. We assume a Kolmogorov spectrum in the inertial range by taking $\nu=5/3$. From the requirement that $\langle b_2^2 \rangle = 2 \pi \int_0^\infty S_2(k) k~dk$, we get
\begin{equation}\label{eq:c2}
C_2 = \frac{(\nu-1)(p+2)}{2 \pi (p + \nu +1)}.
\end{equation}

Note that the inertial range ($k_\perp > 1/\lambda_2$) behavior is described by a conventional power law, and $p$ only determines the long-wavelength properties of the spectrum. The spectral behavior of interplanetary magnetic fluctuations at long wavelengths is not well determined from single point measurements \citep{matthaeus2016prl}, and there are ambiguities surrounding the question of whether the observed structures are spatial or temporal in origin. The observations of ``$1/f$" noise at low frequencies also complicate matters \citep{matthaeus1986prl}. All values of $p \geq -1$ yield power spectra that give rise to a finite energy, but these spectra may be differentiated based on the characteristic length scales associated with them. In addition to the standard correlation scale \citep{Batchelor1953book}, there is a distinct scale, called the ultrascale, which is of importance in applications of 2-D turbulence (\citealt{matthaeus2007spectral} and references therein). The ultrascale is so named because it is generally larger than the correlation scale, and it may be interpreted as  a typical size of an ``island" of 2-D turbulence \citep{matthaeus1999conf} and as the perpendicular coherence length of the FLRW \citep{ruffolo2004ApJ}.

We consider the following cases \citep{matthaeus2007spectral}: $p=-1$ (infinite correlation scale and an infinite ultrascale), $p=0$ (finite correlation scale but an infinite ultrascale), and $p \geq 1$ (finite ultrascale and finite correlation scale). The case $p=2$ is of special interest since it corresponds to homogeneous turbulence. Each of the above possibilities is realizable as each yields a finite energy. However, unlike the correlation scale, the values taken by the ultrascale in space and astrophysical plasmas are not well known, and there is a paucity of established methods to measure it (see \citealt{matthaeus2007spectral} for a proposed technique). Therefore, it is of interest to examine the dependence of the diffusion coefficients on $p$. If there is a marked differentiation between the mfps computed for different cases, then observations of the mfps may be used to infer constraints on the ultrascales prevailing in the heliospheric plasma.

To finally obtain an expression for the perpendicular mean free path, we use Equations \eqref{eq:twod1} and \eqref{eq:twod2} in Equation \eqref{eq:rbd} and set $\nu = 5/3$ to get

\newcommand{\defa}{\frac{v}{\lambda_\parallel \sqrt{2 \langle \tilde{v}^2_x \rangle}}}

\begin{equation} \label{eq:mfp_perp}
\begin{aligned}
\lambda_\perp &= \mathscr{F}_1 \Bigg\{ \frac{\lambda_{2}^{-2/3}}{5 \mathscr{F}_2^{5/3} \sqrt{\pi}} \bigg[ 3 \sqrt{\pi} \mathscr{F}_2^{5/3} \lambda_2^{5/3} \text{Erfc}\left( \mathscr{F}_2 \lambda_2 \right)   \\
             &+ \Gamma\left(\frac{1}{3}\right)- 3 \Gamma\left(\frac{4}{3},\mathscr{F}_2^2 \lambda_2^2 \right)  \bigg]    \\ 
             &+ \delta_{p,-1} \lambda_2 \bigg[ \mathscr{F}_2 \lambda_2 \frac{2}{\sqrt{\pi}} 
\tensor[_2]{\text{F}}{_2} \left(\frac{1}{2},\frac{1}{2};\frac{3}{2},\frac{3}{2};-\mathscr{F}_2^2 \lambda_2^2 \right)
  \\
             &-0.981755 - \log (\mathscr{F}_2 \lambda_2)
\bigg]   \\
             &+(1-\delta_{p,-1})\frac{\lambda_2}{p+1} \bigg[\text{Erfc}(\mathscr{F}_2 \lambda_2) \\
             &- \frac{\mathscr{F}_2 \lambda_2}{\sqrt{\pi}} 
             E_{\frac{p}{2}+1}(\mathscr{F}_2^2 \lambda_2^2) \bigg]
\Bigg\},            
\end{aligned}
\end{equation}
where 
\begin{equation}
\mathscr{F}_1 = \sqrt{\pi^3} C_2 \frac{v \langle b_2^2 \rangle a^2}{B^2 \sqrt{2 \langle \tilde{v}^2_x \rangle}}, 
\end{equation}
and
\begin{equation}
\mathscr{F}_2 = \defa.
\end{equation}
In Equation \eqref{eq:mfp_perp}, Erf is the error function, $\Gamma$ is the gamma function, $\tensor[_2]{\text{F}}{_2}$ is a hypergeometric function, $E_{\frac{p}{2}+1}$ is the generalized exponential integral function, and the Kronecker delta function is used as a switch between the four values of $p$. $C_2$ depends on the value of $p$, as can be seen from Equation~\eqref{eq:c2}. Note that in the corresponding NLGC result \citep{pei2010cosmic}, an implicit method is required to obtain $\lambda_\perp$, in contrast to the RBD result, which is an explicit solution for $\lambda_\perp$.

\section{Solar wind model}

Equations \eqref{eq:mfp_p} and \eqref{eq:mfp_perp} require specification of the large-scale IMF, and the magnetic fluctuation energies and correlation lengths for both slab and 2-D turbulence. For this purpose, we use a Reynolds-averaged Navier-Stokes approach, based on the Reynolds decomposition \citep[e.g.,][]{mccomb1991physics} of a physical field, $\tilde{\mathbf{a}}$, into a mean and a fluctuating component:
\begin{equation}
\tilde{\mathbf{a}} = \mathbf{a}+\mathbf{a'},
\end{equation}
where $\mathbf{a} = \langle \tilde{\mathbf{a}} \rangle$ is an ensemble average, associated with the large scales of motion, and $\mathbf{a'}$ is a fluctuating component, here assumed to be small scale. By construction, $\langle \mathbf{a'} \rangle = 0$. Application of this decomposition to the MHD equations, along with a series of approximations appropriate to the solar wind, leads us to a set of mean-flow equations that are coupled to the small-scale fluctuations via another set of equations for the statistical descriptors of turbulence. For the details of the procedure for handling the fluctuations, we refer the reader to \cite{breech2008turbulence}. 

In this study, we use the solar wind model described in detail by
\cite{usmanov2014three}. It is a global, fully three-dimensional
magnetohydrodynamic model that accounts for the effects of fluctuations in
heating and acceleration of the solar wind flow. The computational domain,
which in the present study extends from the coronal base to 3~AU, is split
into three regions: inner ($1 - 20~R_\odot$), intermediate ($20 - 45~R_\odot$), and outer ($45~R_\odot - 3$ AU). In the inner region, steady-state solutions of one-fluid,
polytropic ($\gamma = 1.08$) solar wind equations with WKB Alfv\'en waves
are obtained by time relaxation starting from an initial state composed of
a Parker-type flow in a dipole magnetic field \citep{usmanov2000global,
usmanov2003tilted}. Two-fluid steady-state equations for protons
and electrons with Hollweg's electron heat flux and WKB Alfv\'en waves are
solved in the intermediate region by forward integration along the radius $r$
\citep{pizzo1978three,pizzo1982three,usmanov1993global}. The boundary conditions for
the intermediate region are extracted from the inner region solution. In
the outer region, we solve three-fluid (thermal protons, electrons, and
pickup protons) Reynolds averaged solar wind equations simultaneously with
transport equations for turbulence energy, cross helicity and
correlation length. Steady-state solutions in the outer region are obtained
by time relaxation, using an eddy-viscosity approximation for the Reynolds stress
tensor and turbulent electric field, with boundary conditions provided
by solutions in the intermediate region \citep{usmanov2014three}. The use of steady-state simulations is justified here since ambient solar wind conditions change on time scales long compared to the time energetic particles spend in the inner heliosphere.

In our calculations, we have used the same input parameters at the coronal
base as in \cite{usmanov2014three}: the driving amplitude of Alfven waves
is set to 35~km~s$^{-1}$, the initial density is $0.4\times
10^8$~cm$^{-3}$, and the initial plasma temperature is $1.8\times 10^6$ K.
The magnetic field magnitude is assigned as the field strength of the
source magnetic dipole on the poles. This parameter is set to 16~G to match the
magnitude of the heliospheric magnetic field observed by Ulysses. The
computations are carried out on a composite spherical grid
\citep{usmanov1996global,usmanov2012three} using the Central Weighted
Essentially Non-Oscillatory (CWENO) spatially third-order reconstruction
algorithm of \cite{kurganov2000third}.  The spatial CWENO
discretization is combined with the Strong Stability-Preserving Runge-Kutta
scheme of \cite{gottlieb2001strong} for time integration and the method of
\cite{powell1994approximate} for maintaining the $\nabla\cdot{\bf B} = 0$ condition.

For our purposes here, we extract from the outer region simulation ($45~R_\odot - 3$ AU) the mean magnetic field, $\mathbf{B}$, the fluctuation energy, $Z^2$ (defined below), and the correlation length for the turbulence, $\lambda$. Here,
\begin{equation} \label{eq:Z}
Z^2 = \langle v'^2 + b'^2 \rangle,
\end{equation}
is twice the turbulent energy per unit mass, defined in terms of the velocity and magnetic field fluctuations, $\mathbf{v}'$ and $\mathbf{B}'$, respectively. The amplitude of magnetic fluctuations has been normalized to Alfv\'{e}n units using $\mathbf{b}' = \mathbf{B}'(4 \pi \rho)^{-1/2}$, where $\rho$ is the mass density. To extend our calculation closer to the sun, we use data from the inner ($1 - 20~R_\odot$) and intermediate ($20 - 45~R_\odot$) regions, where the simulation does not have a turbulence model for $Z^2$ and $\lambda$. Here we use the the WKB Alfv\'{e}n wave energy density \citep{usmanov2000global}, $\mathcal{E}$, as a proxy for the turbulent fluctuation energy via $Z^2 = 2 \mathcal{E} /\rho$. To get an approximation for the correlation scale in these regions, we use the hypothesis from \cite{Hollweg1986JGR} that the correlation length varies as the distance between magnetic field lines, which in turn depends on the field strength \citep{spruit1981NASSP}, so that $\lambda \propto B^{-1/2}$. We set the constant of proportionality such that $\lambda$ at the boundaries of the intermediate and outer regions matches. We are currently working on refinements of the model that will modify the region in which turbulence modeling is included, so that this region will extend closer to the sun. 

To proceed with the calculation of the mfps, some assumptions must be made in order to relate the correlation scale of our turbulence model ($\lambda$) to the slab and 2-D correlation scales in Equations~\eqref{eq:mfp_p} and \eqref{eq:mfp_perp}, respectively. First, we note that the turbulent fluctuations in our model are primarily transverse to the mean magnetic field \citep{breech2008turbulence}, and thus identify the correlation scale of 2-D turbulence to be equal to the correlation scale of our turbulence model, so that $\lambda_2 = \lambda$. Observational studies \citep{osman2007ApJ654,Weygand2009JGRA,weygand2011JGR116} indicate that the slab correlation scale is about a factor of two larger than the 2-D correlation scale, and accordingly, we assume $\lambda_s = 2 \lambda_2$. In our approximate treatment, we assume in effect that the magnetic and velocity correlation functions are structurally similar \citep{zank1996evolution}, so that the magnetic correlation length is found to be equal to the single correlation scale that we follow dynamically. In the inner heliosphere where the cross helicity is large, it becomes advantageous to employ a two correlation length theory \citep{mattheus1994JGR99,wan2012JFM697,Zank2012ApJ745,Zank2017ApJ835}, as has been implemented, e.g., by \cite{Adhikari2015ApJ805}.

To approximate the energy in slab and 2-D magnetic fluctuations, we first convert $Z^2$ to $B'^2$ using Equation~\eqref{eq:Z}:
\begin{equation}\label{eq:mag_fluc}
\langle B'^2 \rangle = \frac{Z^2}{r_A+1} 4 \pi \rho,
\end{equation}
where $r_A = \langle v'^2 \rangle /\langle b'^2 \rangle$ is the Alfv\'{e}n ratio. An accurate dynamical model for $r_A$ is desirable, but must include complications such as non-local effects \citep[e.g.,][]{grappin1983AaA26,mattheus1994JGR99,hossain1995PhFl}. At present we maintain a simpler approach, and take $r_A$ to have a value of 1 in the inner and intermediate regions ($1 - 45~R_\odot$), and  a value of $1/2$ for heliocentric distances larger than $45~R_\odot$. These values are motivated by spacecraft observations \citep{Tu1995SSRv}, but we recognize that attempts have been made to treat $r_A$ dynamically \citep{grappin1983AaA26,marsch1989JPlPh41,tu1990JPlPh44,mattheus1994JGR99,yokoi2007PhPl,Zank2012ApJ745}. See especially the comparison with observations by \cite{Adhikari2015ApJ805} and \cite{Zank2017ApJ835}.

Next, recalling the assumption that the magnetic fluctuations have a dominant 2-D component with a small slab contribution, and following observations \citep{matthaeus1990JGR,bieber1994proton} that find the ratio of the 2-D and slab energies to be 80\% to 20\%, we use
\begin{equation}\label{eq:slab_2d}
\frac{\langle b^2_s \rangle}{\langle b^2_2 \rangle} = \frac{20}{80} = 0.25
\end{equation} 
to compute the slab and 2-D fluctuation energies from Equation~\eqref{eq:mag_fluc} and  $\langle b^2_2 \rangle + \langle b^2_s \rangle = \langle B'^2 \rangle$. In recent work by \cite{Hunana2010ApJ718} and \cite{Zank2017ApJ835}, refinements to this simplified perspective on the breakdown of the slab and 2-D fluctuation energies are discussed. In particular, \cite{Zank2017ApJ835} solve separate equations for the slab and 2-D energies with a simplified IMF and background solar wind flow. They find that the evolution of the two components is markedly different in the outer heliosphere (beyond $\sim3$ AU), where driving by pickup ions leads to an increase in the slab component's energy, while the energy of the 2-D component continues to decrease with heliocentric distance. Their results show, however, that the radial evolution of slab and 2-D energies is not too dissimilar below 3 AU. Similar results are presented by \cite{Oughton2011JGRA116} using their two-component model. Therefore, for the purposes of our present work, where we focus on the inner heliosphere, our simple decomposition of $\langle B'^2 \rangle$ into slab and 2-D components, using the constant ratio expressed in Equation~\eqref{eq:slab_2d}, seems appropriate. Studies of CR diffusion in the outer heliosphere would undoubtedly benefit from using a two-component turbulence transport model. A detailed assessment of different transport equations for turbulence is beyond the scope of this work.

\section{Results}

\subsection{Solar wind  model results}
\begin{figure}
\includegraphics[scale=.37]{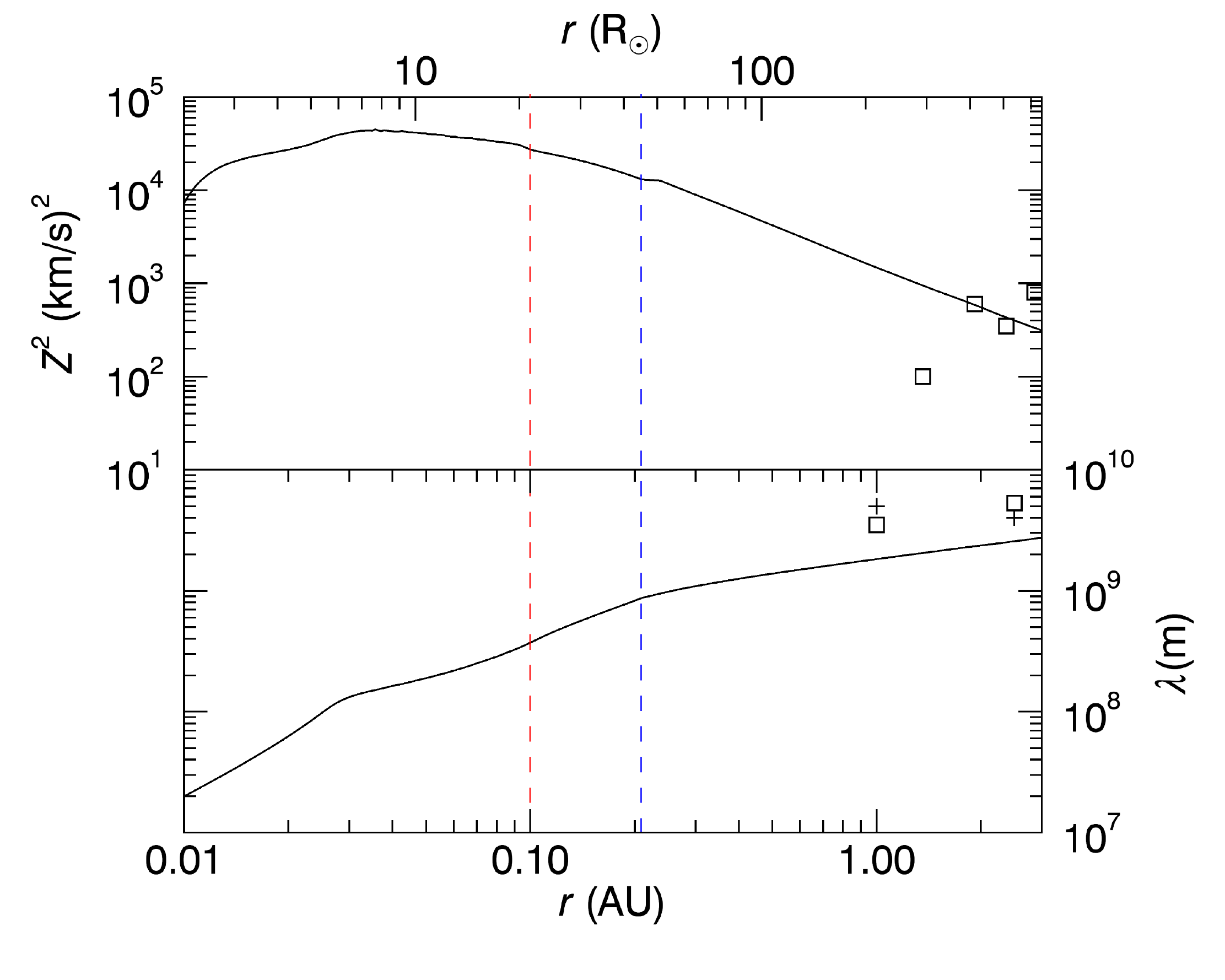}
\caption{Model results near the ecliptic plane, for a run with an untilted solar dipole, are compared with observational data from Voyager 2, Helios, and the NSSDC Omnitape. The $Z^2$ data are from \cite{zank1996evolution}, and the $\lambda$ data are from \cite{smith2001JGR}. The solid lines are from our simulations. The different symbols represent different methods of calculation. The dashed vertical lines represent the boundaries of the different simulation regions, with red marking the inner-intermediate region boundary at $20~R_\odot$, and blue marking the intermediate-outer region boundary at $45~R_\odot$, respectively. Note that the observations are for various times in the solar cycle, and are shown here for general context only.}
\label{fig:ingred1}
\end{figure}

We begin our presentation of the results with a discussion of the core fields from the simulation - $B, \lambda$, and $Z^2$ - which are the ingredients that go into our calculation of the diffusion coefficients. Figure \ref{fig:ingred1} shows the radial evolution of the turbulence energy and the turbulence correlation scale from our model and simulation with an untilted dipole source. The data are for a $7\degree$ heliolatitude, which we take to be the broadly-defined ecliptic region. Also shown are observational results from Voyager 2, Helios, and the National Space Science Data Center (NSSDC) Omnitape dataset, indicating a reasonable agreement with the simulation results. The observational data for $Z^2$ and $\lambda$ are from \cite{zank1996evolution} and \cite{smith2001JGR}, respectively. Note that the observations are for various times in the solar cycle, and are shown here for general context only. The dashed vertical lines in Figure~\ref{fig:ingred1} represent the boundaries of the different simulation regions, with red marking the inner-intermediate region boundary at $20~R_\odot$, and blue marking the intermediate-outer region boundary at $45~R_\odot$, respectively. Note that we present results for $r>2~R_\odot$ ($r$ is the radial distance measured from the solar center), even though the inner boundary of the inner region simulation is at $1~R_\odot$. The parallel mfp acquires extremely large values ($> 10$ AU) in the region very close to the solar surface, due to the large values of $B$ prevailing there. These large values of $\lambda_\parallel$ are not of physical relevance and present problems for visualization, and we therefore restrict our results to $r>2~R_\odot$.

\begin{figure} 
\includegraphics[scale=.4]{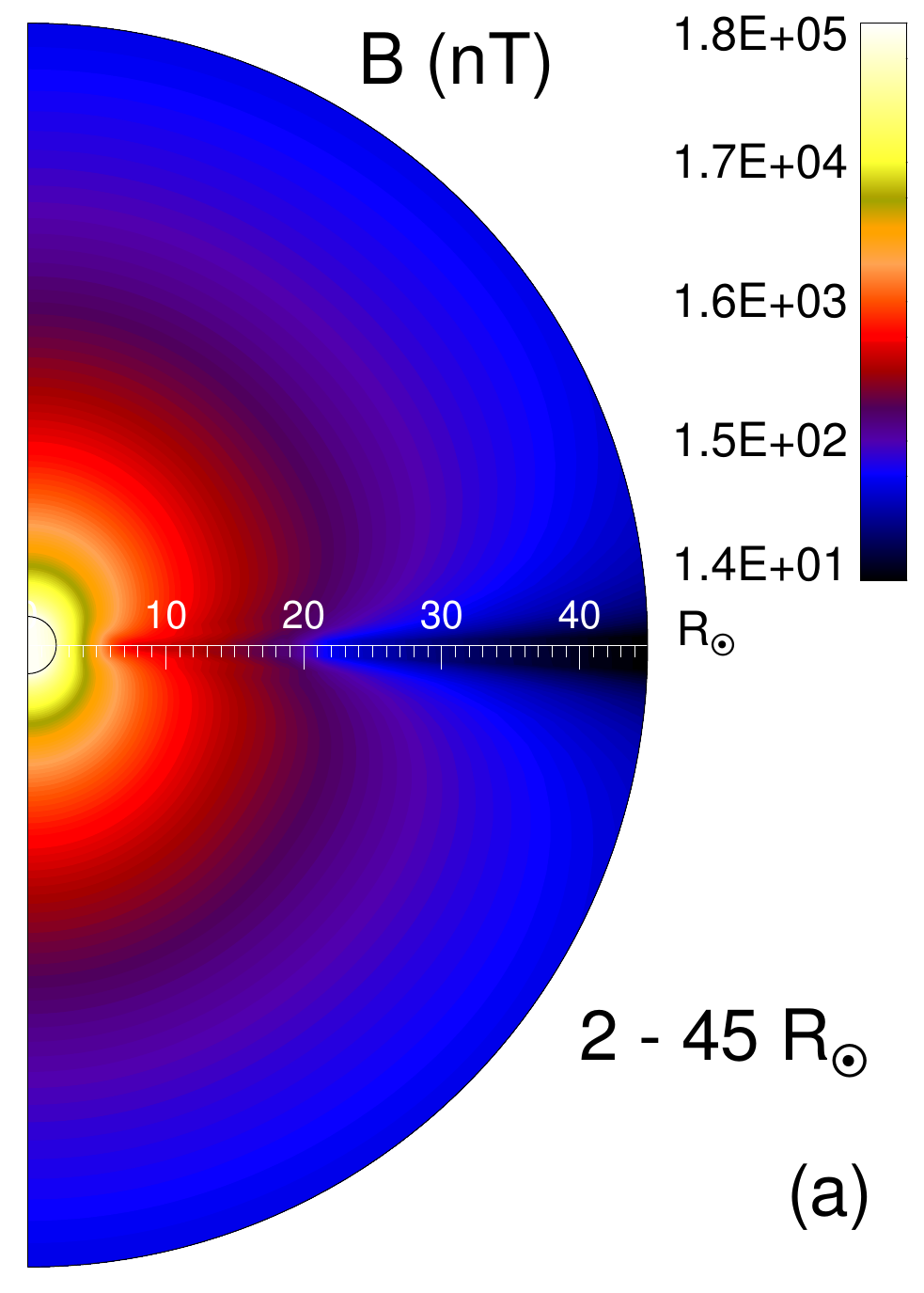}
\includegraphics[scale=.4]{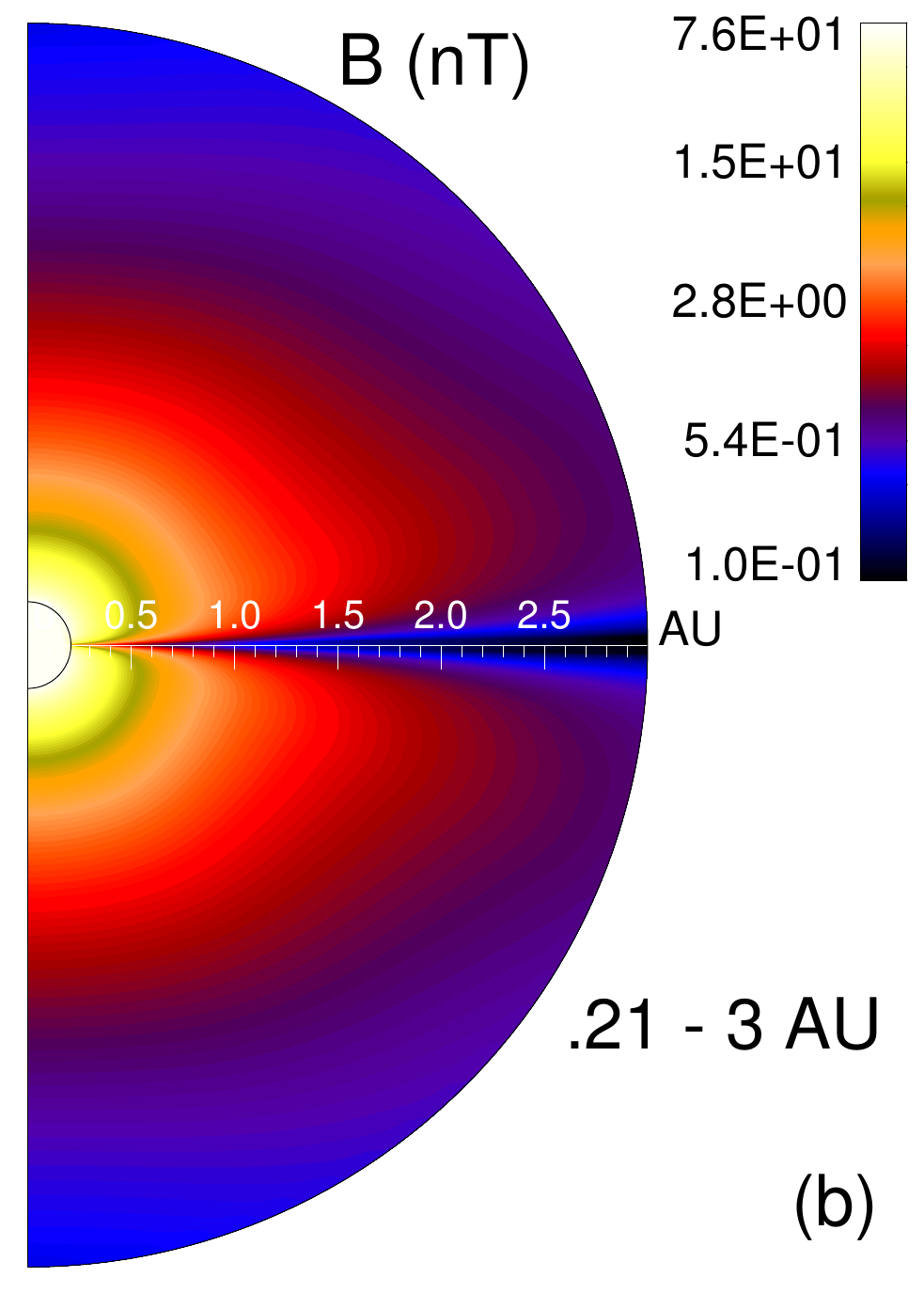}
\includegraphics[scale=.4]{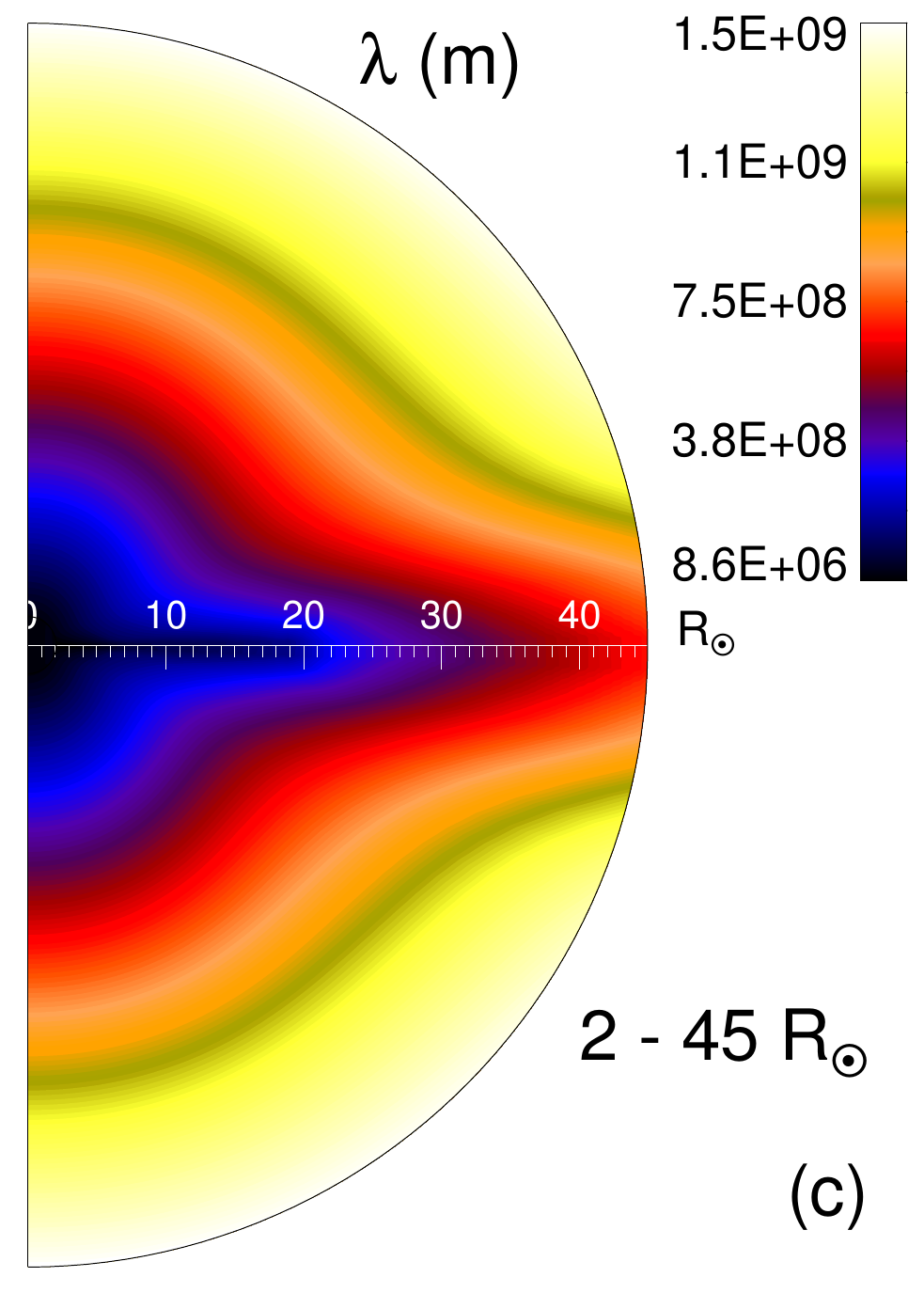}
\includegraphics[scale=.4]{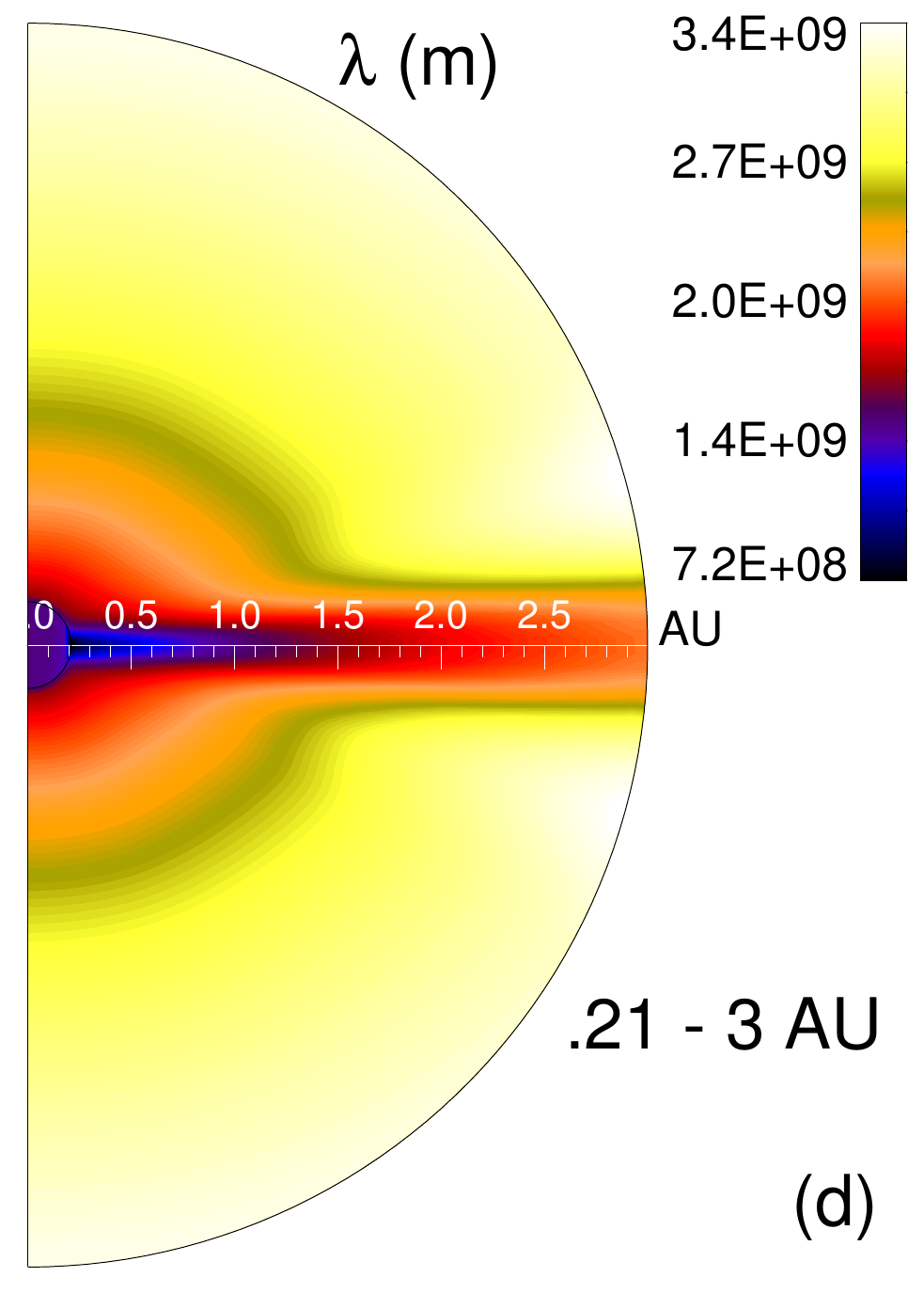}
\includegraphics[scale=.4]{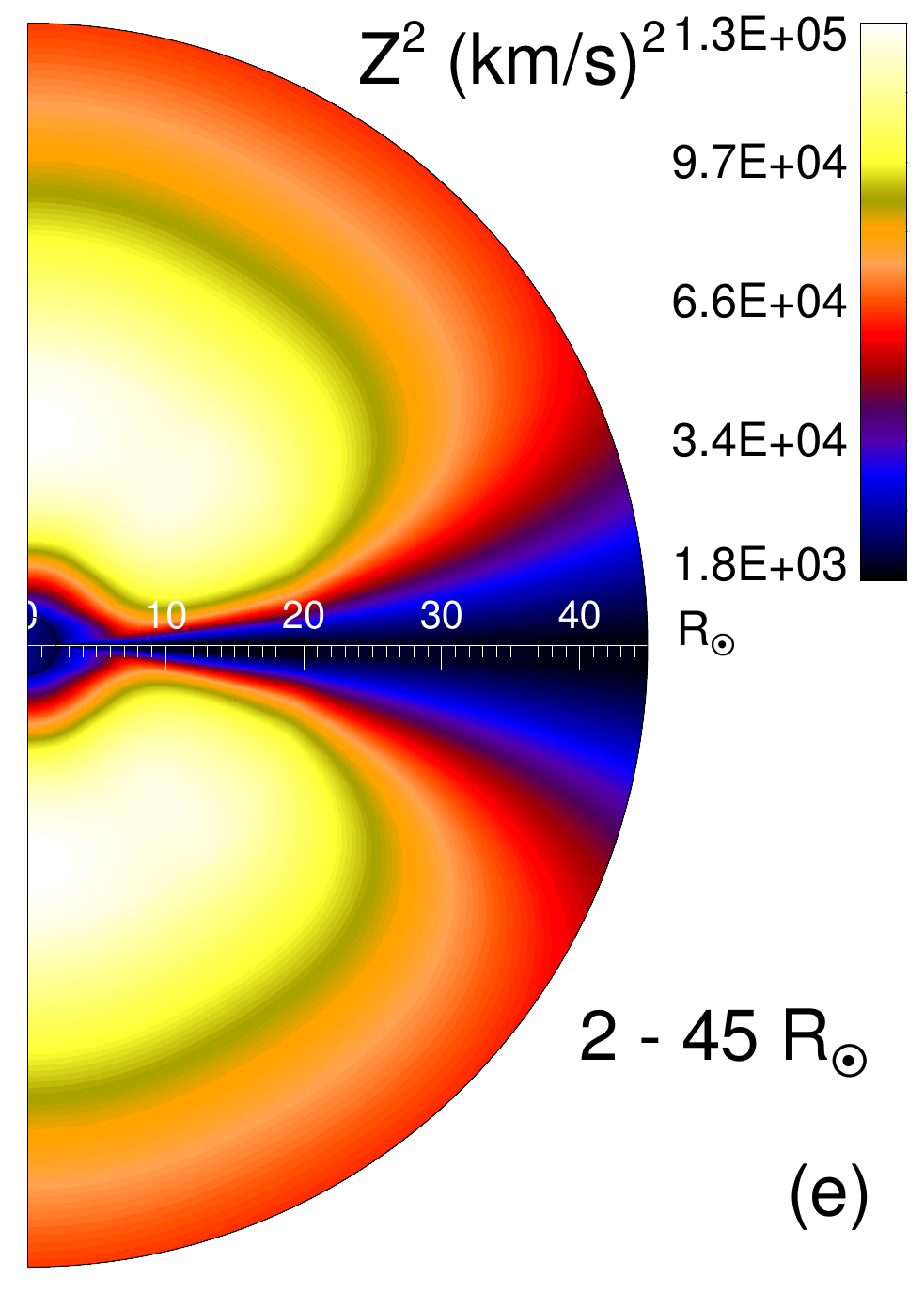}
\includegraphics[scale=.4]{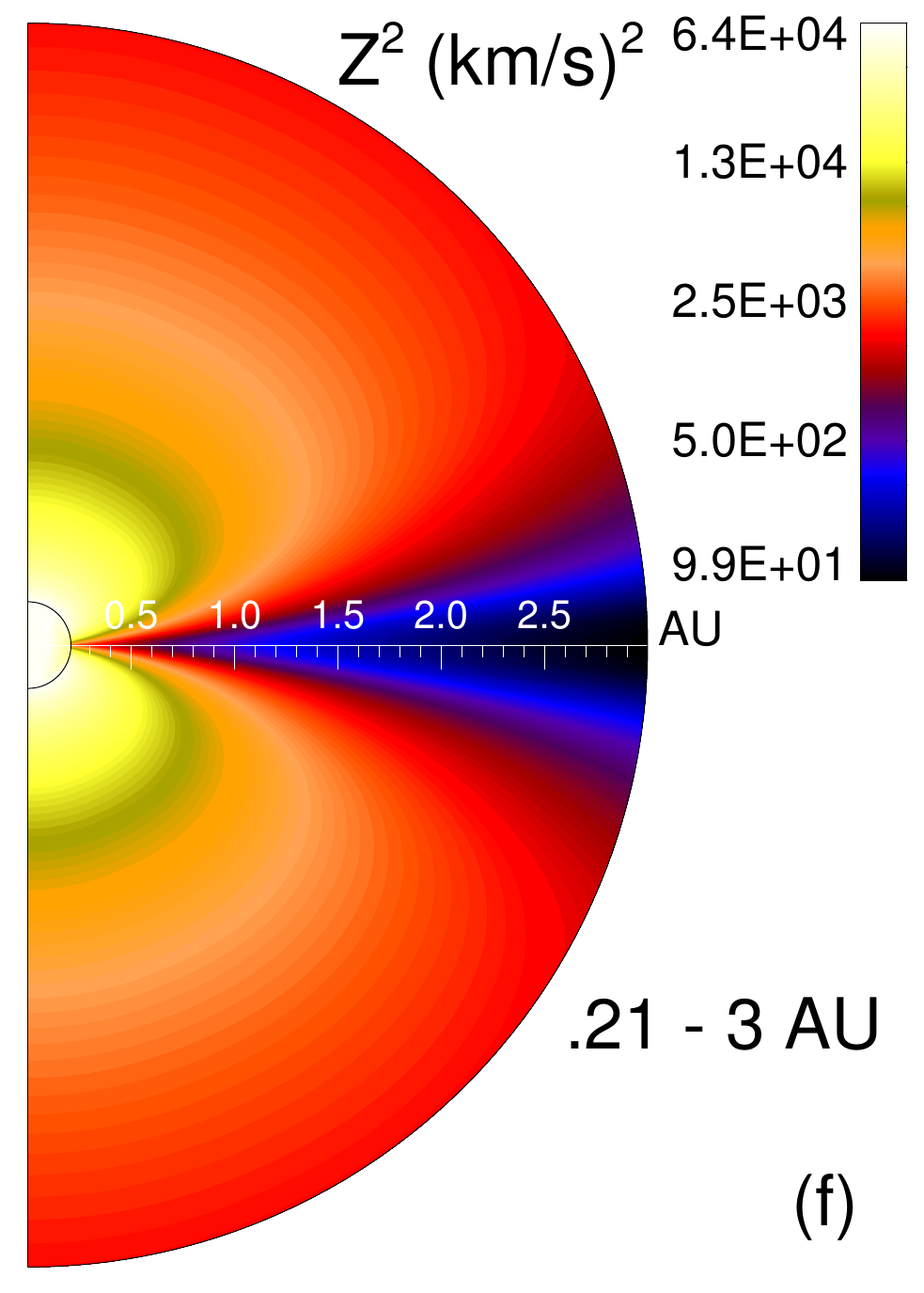}
\caption{Contour plots of the heliospheric magnetic field ($B$), the turbulence correlation scale ($\lambda$), and the turbulence energy ($Z^2$) in the meridional plane for an untilted solar dipole. The figures on the left cover $2 - 45~R_\odot$, and the ones on the right cover $0.21 - 3$ AU ($45 - 645~R_\odot$).}
\label{fig:ingred2}
\end{figure}

Figure \ref{fig:ingred2} shows the distribution in the meridional plane of the three ingredients - $B, Z^2,$ and $\lambda$ - for a simulation with an untilted source dipole. The figures on the left are from the inner and  intermediate regions ($2 - 45~R_\odot$), and the ones on the right are from the outer region ($0.21 - 3$ AU). For a detailed discussion of these simulation results, we refer the reader to \cite{usmanov2000global} and \cite{usmanov2014three}. We note here that the magnetic field results agree well with Ulysses observations \citep[see Figure~8 of][]{usmanov2014three}, with the field vanishing at the heliospheric current sheet (HCS) at $0\degree$ heliolatitude. The turbulence correlation scale increases with heliocentric distance, as is well known from observations \citep{Tu1995SSRv}. The turbulence energy increases on moving from the ecliptic plane towards higher heliolatitudes because of shear interactions between slow (low latitude) and fast (high latitude) wind \citep[See, e.g.,][]{breech2008turbulence}. In the following subsections, we will discuss how these distributions influence the behaviour of the diffusion length-scales.

\subsection{Radial evolution of mean free paths}

\begin{figure}
\includegraphics[scale=.36]{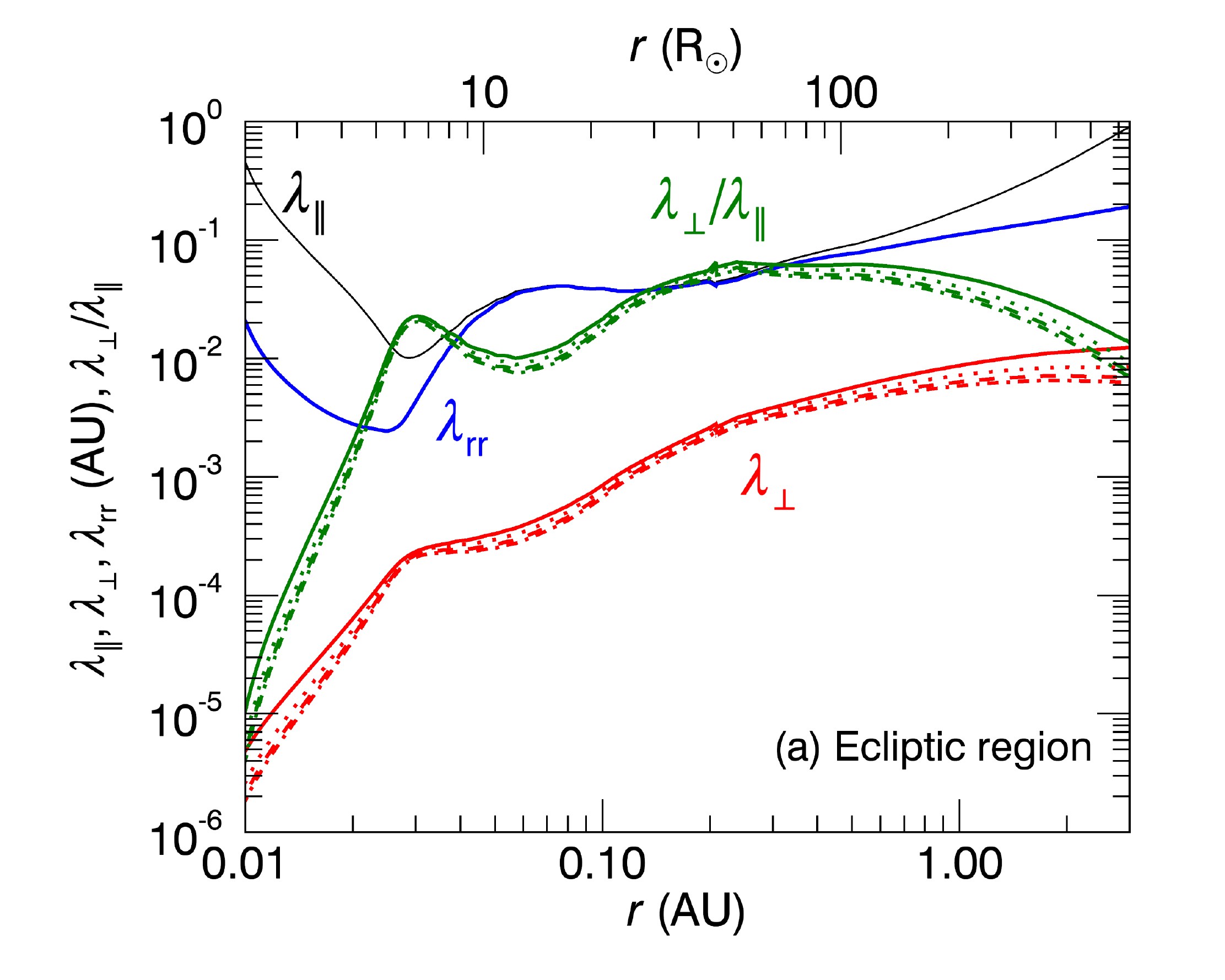}
\includegraphics[scale=.36]{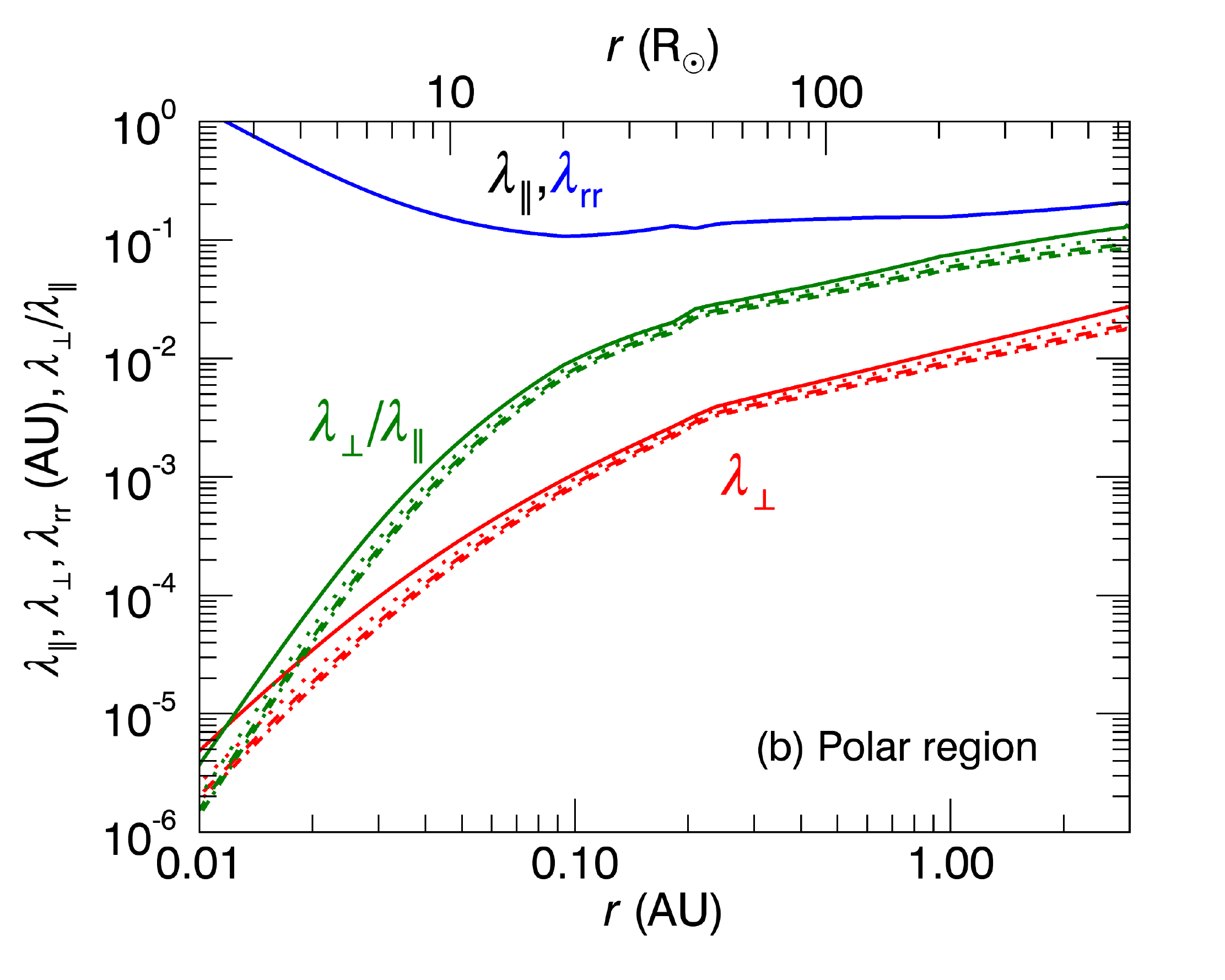}
\caption{Radial dependence of the parallel (black), perpendicular (red), and radial (blue) mfps (a) near the ecliptic plane ($7\degree$ heliolatitude) and (b) near the pole ($86\degree$ heliolatitude). Also shown is $\lambda_\perp/\lambda_\parallel$ (green). The solid lines are for $p=-1$, the dotted lines for $p=0$, the dashed lines for $p=1$, and the dash-dotted lines for $p=2$. Proton rigidity is 445 MV (100 MeV kinetic energy). Note that the curves for $\lambda_\parallel$ and $\lambda_{rr}$ coincide in (b).}
\label{fig:mfp_untilt}
\end{figure}

In Figure \ref{fig:mfp_untilt} we show the radial evolution of the parallel, perpendicular, and radial mfps (black, red, and blue lines, respectively) in the ecliptic region (Figure \ref{fig:mfp_untilt}a) and near the solar rotation axis ($86\degree$ heliolatitude, Figure \ref{fig:mfp_untilt}b), for an untilted source dipole. Also shown is the ratio of the perpendicular mfp to the parallel mfp (green lines). The solid, dotted, dashed, and dash-dotted lines correspond to $p=-1,0,1,$ and 2, respectively, and the mfps are computed for protons with rigidity equal to 445 MV, corresponding to a kinetic energy of 100 MeV. Here we would like to remind the reader that our turbulence parameters ($Z^2$ and $\lambda$) in the region $1 - 45~R_\odot$ are not from the turbulence model, but are calculated using the approximations detailed in Section 3. As such, these results represent a preliminary attempt at mapping the diffusion length scales in a region that will soon be investigated by upcoming spacecraft missions such as Solar Probe Plus. 

Near the ecliptic plane (Figure \ref{fig:mfp_untilt}a), as one moves outward from the solar surface, the increasing strength of the turbulence energy (see Figure \ref{fig:ingred1}) leads to a sharp decrease in $\lambda_\parallel$  in the region $2 - 5~R_\odot$, with the rapidly decreasing IMF reinforcing this behaviour. In this region, $\lambda_\parallel \propto r^{-3.46}$, and there is a corresponding increase in $\lambda_\perp (\propto r^{3.55} \text{ for } p=-1 \text{ and } \propto r^{4.34} \text{ for } p=2)$. Since the IMF has a significant meridional component here, the large winding angle ($\Psi$) between the radial direction and the IMF leads to $\lambda_\perp$ having an influence on the radial mfp (see Equation \ref{kappa_r}), with $\lambda_{rr} \propto r^{-1.97}$. From $0.03 - 3$ AU, $\lambda_\parallel$ mostly increases as $r^{0.82}$, and $\lambda_\perp$ as $r^{0.79}$. From 0.1 to 3 AU, $\Psi$ is once again large because of the increased azimuthal component of the IMF, and $\lambda_\perp$ reduces the radial mfp, with $\lambda_{rr} \propto r^{0.53}$. Observational studies for $r< 3$ AU have found $\lambda_{rr} \propto r^b$ with $b$ ranging from $0.4 - 0.7$ \citep{beeck1987ApJ322}. Note that the radial mfp depends on the value of $p$ (through $\lambda_\perp$), but the $\lambda_{rr}$ curves for different $p$ coincide.

Moving on to the radial evolution of the mfps in the polar region, Figure \ref{fig:mfp_untilt}b shows that the radial mfp is completely dominated by $\lambda_\parallel$. This is because the IMF is near radial at the poles, with a very small winding angle. At the poles, $\lambda_{rr} \propto r^{-1.1}$ until 0.1 AU, after which it remains nearly constant, with identical behavior exhibited by $\lambda_\parallel$. From $2~R_\odot - 0.2$ AU, $\lambda_\perp \propto r^{2.10}$ for $p=-1$ and $\lambda_\perp \propto r^{2.34}$ for $p=2$. From $0.2 - 3$ AU, $\lambda_\perp \propto r^{0.78}$ for $p=-1$ and $\lambda_\perp \propto r^{0.69}$ for $p=2$.

\begin{figure}
\includegraphics[scale=.36]{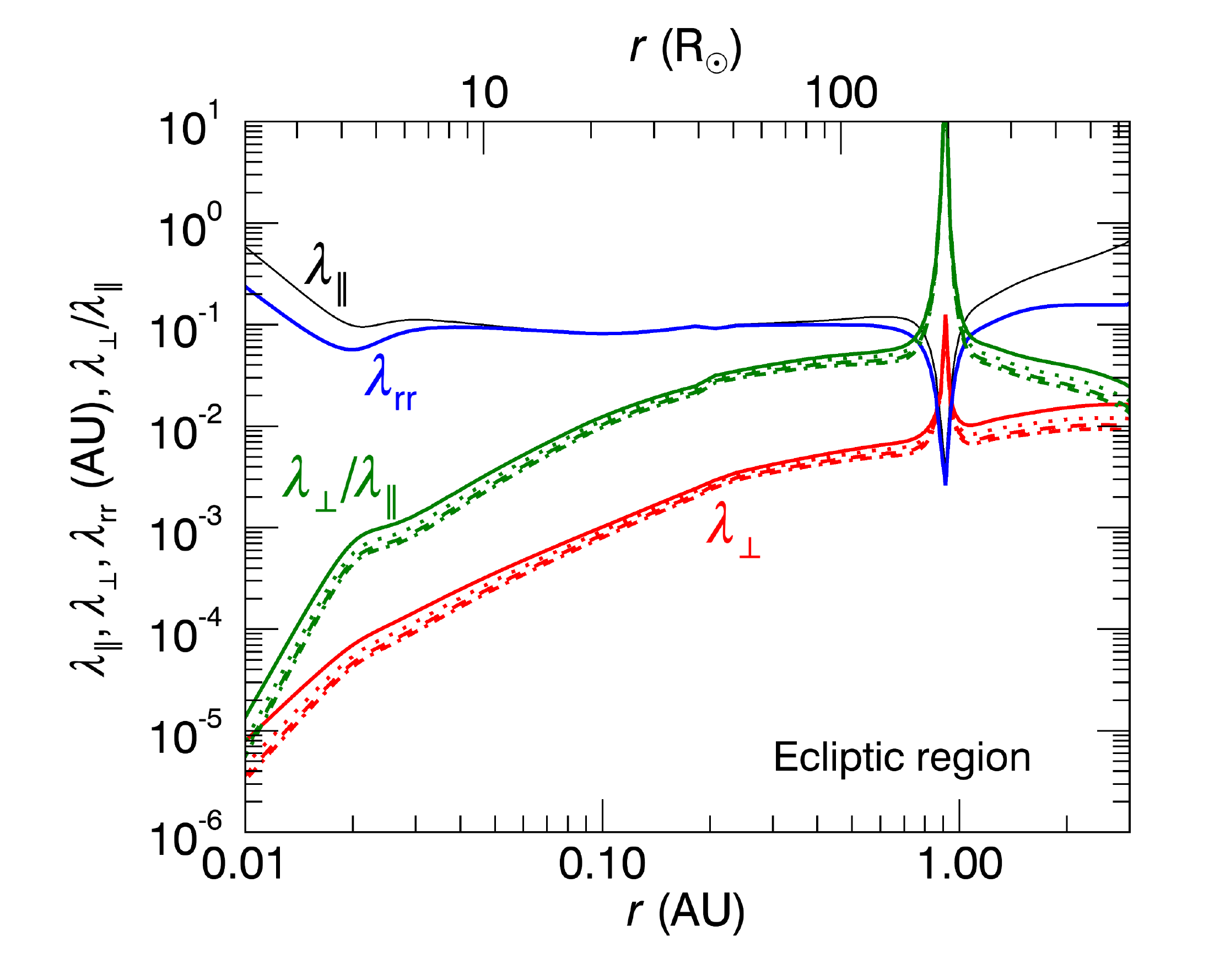}
\caption{Radial dependence of the parallel (black), perpendicular (red), and radial (blue) mfps near the ecliptic plane ($7\degree$ heliolatitude), with a solar dipole having a $30\degree$ tilt. For our particular choice of azimuthal angle ($26\degree$), an HCS crossing occurs at 0.8 AU. Also shown is $\lambda_\perp/\lambda_\parallel$ (green). The solid lines are for $p=-1$, the dotted lines for $p=0$, the dashed lines for $p=1$, and the dash-dotted lines for $p=2$. Proton rigidity is 445 MV (100 MeV kinetic energy).}
\label{fig:mfp_tilt}
\end{figure}

Figure \ref{fig:mfp_tilt} shows the effect of a source dipole with a $30\degree$ tilt when one encounters the heliospheric current sheet (HCS) at around 1 AU: $\lambda_\parallel$ goes through a sudden dip of almost two orders of magnitude, while $\lambda_\perp$ has a corresponding increase of around an order of magnitude. (The radius where the HCS crosses our chosen heliolatitude of $7\degree$ depends on our choice of the azimuthal angle for which we plot results as a function of radius.) The vanishing mean magnetic field and non-vanishing turbulence amplitude at the HCS explain this behaviour, which will be further illustrated in the next subsection discussing the 2-D variation of the mfps in the meridional plane. We note from Figures \ref{fig:mfp_untilt} and \ref{fig:mfp_tilt} that the ratio $\lambda_\perp/\lambda_\parallel$ stays between 0.1 and 0.01 for most of the inner heliosphere, but it exceeds unity at the HCS. Keeping in mind that the current sheet is a singular region in our simulation, in its vicinity the fields do possess physically realizable values. Therefore we may stress the fact that similarly large values of $\lambda_\perp/\lambda_\parallel$ have been observed \citep{dwyer1997ApJ,zhang2003ApJ}. We will come across these domains of significant perpendicular diffusion once again in the meridional plane contours in Section 4.5, below.

In the results presented so far the choice of the long wavelength spectral index $p$ does not significantly alter the mfps, with $\lambda_\perp$ for $p=-1$ generally not more than a factor of two larger than $\lambda_\perp$ for $p=2$. Referring to the discussion in Section 2.2, this result indicates a rather weak dependence of the mfps on the ultrascale (via different $p$ values). The exception appears very close to the solar surface ($2~R_\odot$) in Figure \ref{fig:mfp_untilt}, where the perpendicular mean free path for the $p=-1$ case is several times larger than that for the $p=2$ case. This behaviour may be probed further in simulations with improved coronal turbulence models that are more reliable at such small heliocentric distances. In the following results, unless specified otherwise, we will choose $p=2$, which corresponds to homogeneous turbulence.

\begin{figure}
\includegraphics[scale=.36]{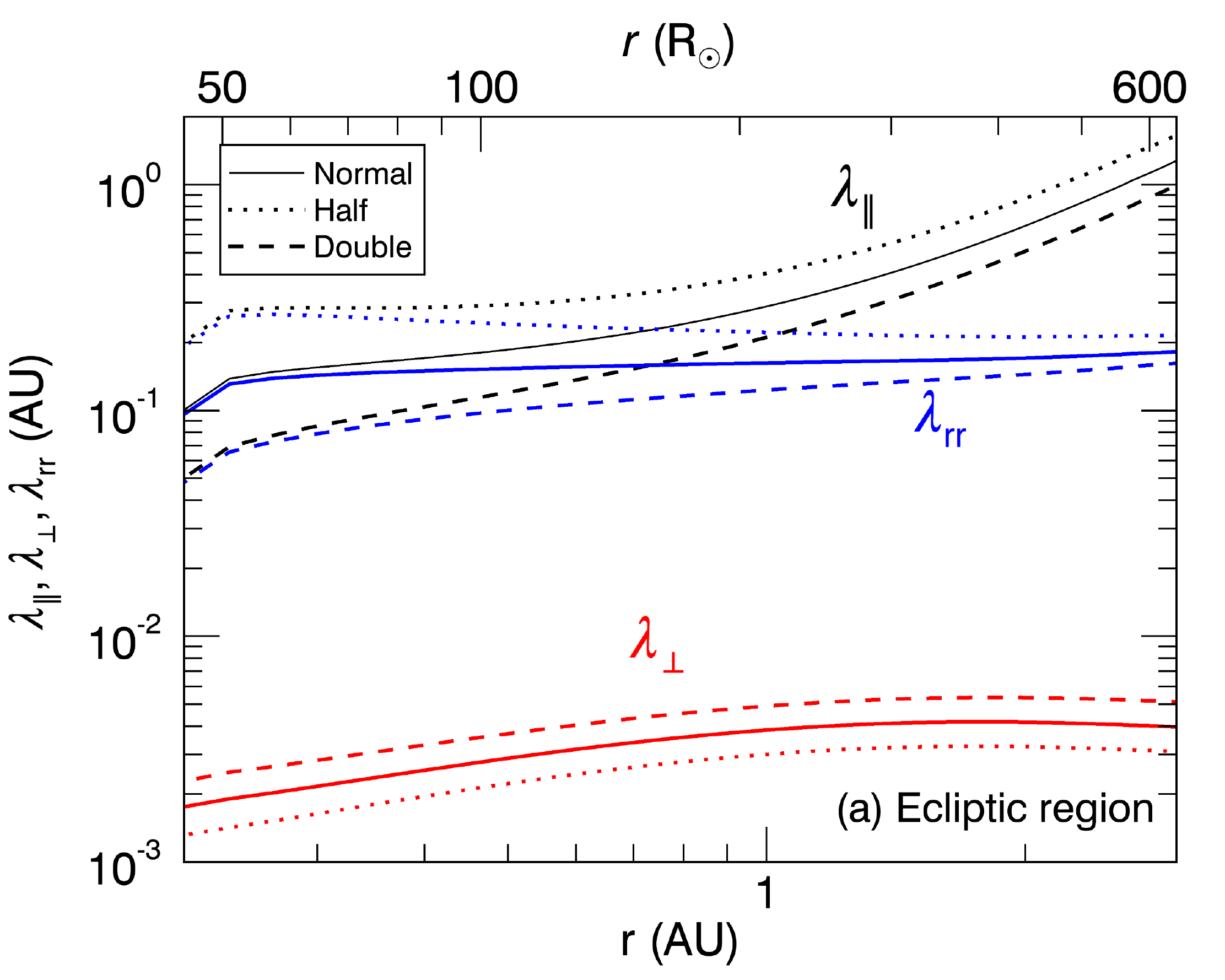}
\includegraphics[scale=.36]{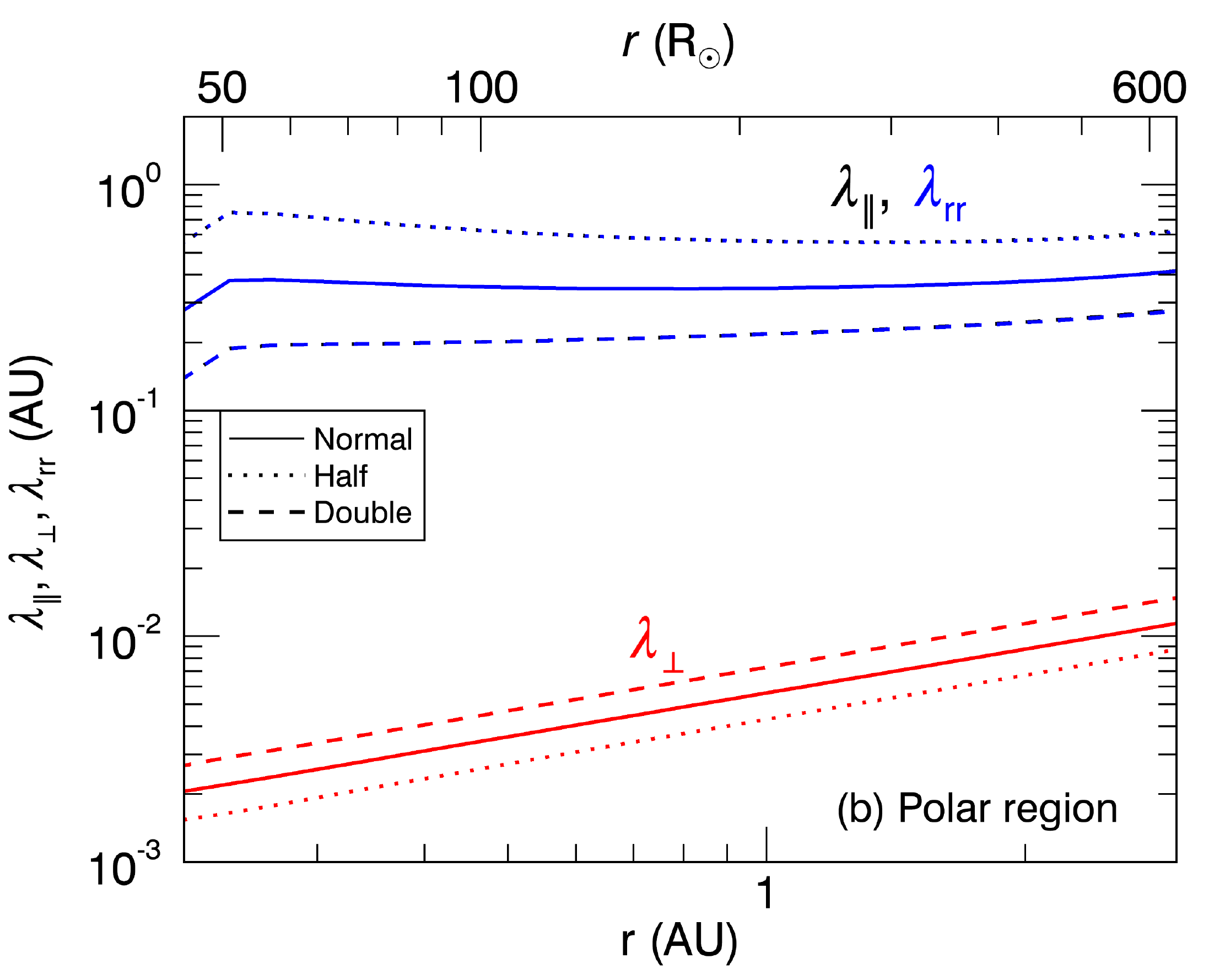}
\caption{Radial dependence of the parallel (black), perpendicular (red), and radial (blue) mfps (a) near the ecliptic plane ($7\degree$ heliolatitude) and (b) in the  polar region ($86\degree$), for varying turbulence amplitudes, with $p=2$. The dashed and dotted lines represent simulations with the turbulence energy ($Z^2$) at the inner boundary of the outer region ($45~R_\odot$) doubled and halved, respectively, relative to a standard level. See text for more details. Note that the curves for $\lambda_\parallel$ and $\lambda_{rr}$ coincide in (b).}
\label{fig:turb_var}
\end{figure}

In Figure \ref{fig:turb_var} we examine the effect of varying the turbulence energy amplitude at the inner boundary (45 $R_\odot$) of the outer region of the simulation, again for 100 MeV protons. Such variation may arise due to solar activity. The solid lines represent a standard $Z^2$ specified at the inner boundary, and dashed and dotted lines represent simulations performed with double and half of this standard value specified at the inner boundary, respectively. In the ecliptic region ($7\degree$ heliolatitude), Figure \ref{fig:turb_var}a indicates, as expected, that an increasing turbulence level leads to a decrease in $\lambda_\parallel$ (and consequently $\lambda_{rr}$). The stronger turbulence increases $\lambda_\perp$ in proportion to $Z$, and therefore increases the extent to which particles may diffusively penetrate the heliosphere. Comparing Figures \ref{fig:turb_var}a and \ref{fig:turb_var}b, it is interesting to note that in the ecliptic region, varying turbulence at the inner boundary leads to an effect on $\lambda_\parallel$ that becomes less pronounced with radial distance. This is not the case in the polar regions with fast wind, however, where the turbulence is less ``aged" compared with low latitudes \citep{matthaeus1998JGR}. Stream interactions near the ecliptic plane reduce the turbulence at a faster rate compared to the rate in the polar regions far from such shearing interactions.

\begin{table}
\setlength{\tabcolsep}{5pt}
\renewcommand{\arraystretch}{1.2} 
\centering
\caption{Parallel mfps in AU for 100 MeV protons at in the ecliptic region at 1 AU. B1, B2, and B3 are from \cite{breech2008turbulence}; P1 and P2 are from \cite{pei2010cosmic}; Cases 1 - 3 are our solutions for varying turbulence levels. Note that our calculation of $\lambda_\parallel$ is independient of $p$.} \label{tab1}
\begin{tabular}{c c c c c c c c c}
$p$ & B1 & B2 & B3 & P1 & P2 & Case 1 & Case 2 & Case 3 \\
\hline
-1  &  2.92 & 6.86 & 1.64 & 0.92 & 0.47 & \multirow{4}{*}{0.29} & \multirow{4}{*}{0.21} & \multirow{4}{*}{0.40} \\
 0   &  2.33 & 5.49 & 1.31 & 0.74 & 0.38 & & & \\
 1   &  2.14 & 5.03 & 1.20 & 0.68 & 0.35 & & & \\
 2   &  2.04 & 4.80 & 1.15 & 0.64 & 0.33 & & & \\
\hline
\end{tabular}
\end{table}

We end this subsection by comparing our solutions in the ecliptic plane with ``consensus" constraints on observations \citep{palmer1982RvGSP,bieber1994proton}. Based on information compiled from several sources, the Palmer consensus finds that for particles in the rigidity range $0.5 - 5000$ MV, $\lambda_\parallel = 0.08 - 0.3$ AU. We note here that the values for the mfps obtained by fitting observational data may depend on the model used; \cite{reames1999SSRv} reviews some such results and suggests a higher parallel mfp of $\sim 1$ AU. Our $\lambda_\parallel$ for a 100 MeV proton at 1 AU varies from $0.29 - 0.40$ AU, and fits the consensus range well. Our solutions are smaller than the values from \cite{breech2008turbulence} and \cite{pei2010cosmic}, which we list in Table \ref{tab1}, along with our results. Here, cases 1, 2, and 3 refer to standard, doubled, and halved turbulence levels, as described above. Note that unlike our calculation of $\lambda_\parallel$, the calculations from \cite{breech2008turbulence} and \cite{pei2010cosmic} depend on the value of $p$. 

Our improved agreement with the Palmer consensus range may be attributed to two improvements in modeling: (1) Here $B$ is a spatially varying field computed dynamically from a self-consistent 3-D  model, in contrast to the Parker-type model used in \cite{breech2008turbulence} and \cite{pei2010cosmic}; (2) The effect of shear interactions is computed self-consistently in our turbulence model \citep{usmanov2014three}, unlike in \cite{breech2008turbulence} and \cite{pei2010cosmic}, where a shear-driving parameter is employed.


\subsection{Latitudinal evolution of mean free paths}

\begin{figure}
\includegraphics[scale=.36]{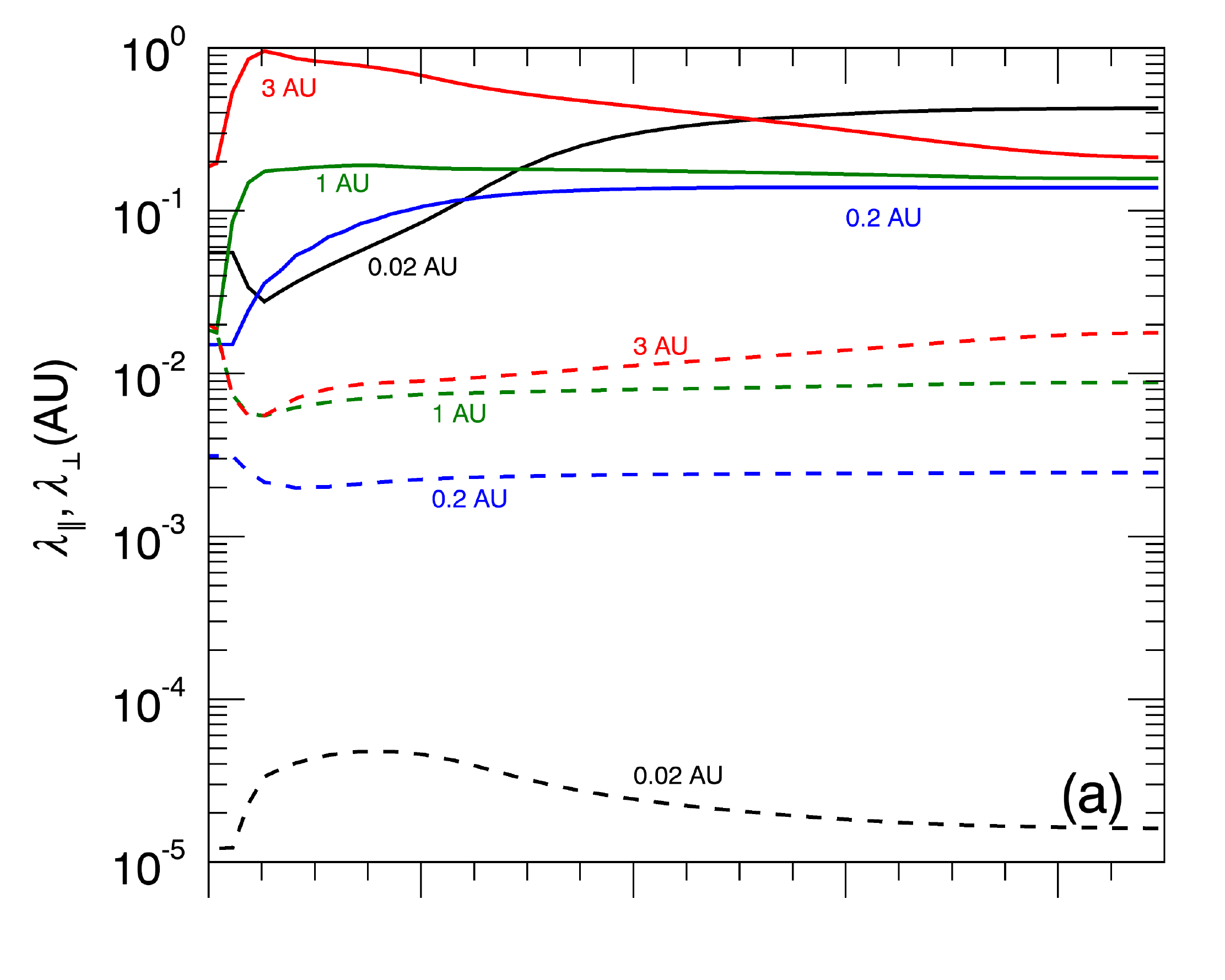}
\includegraphics[scale=.36]{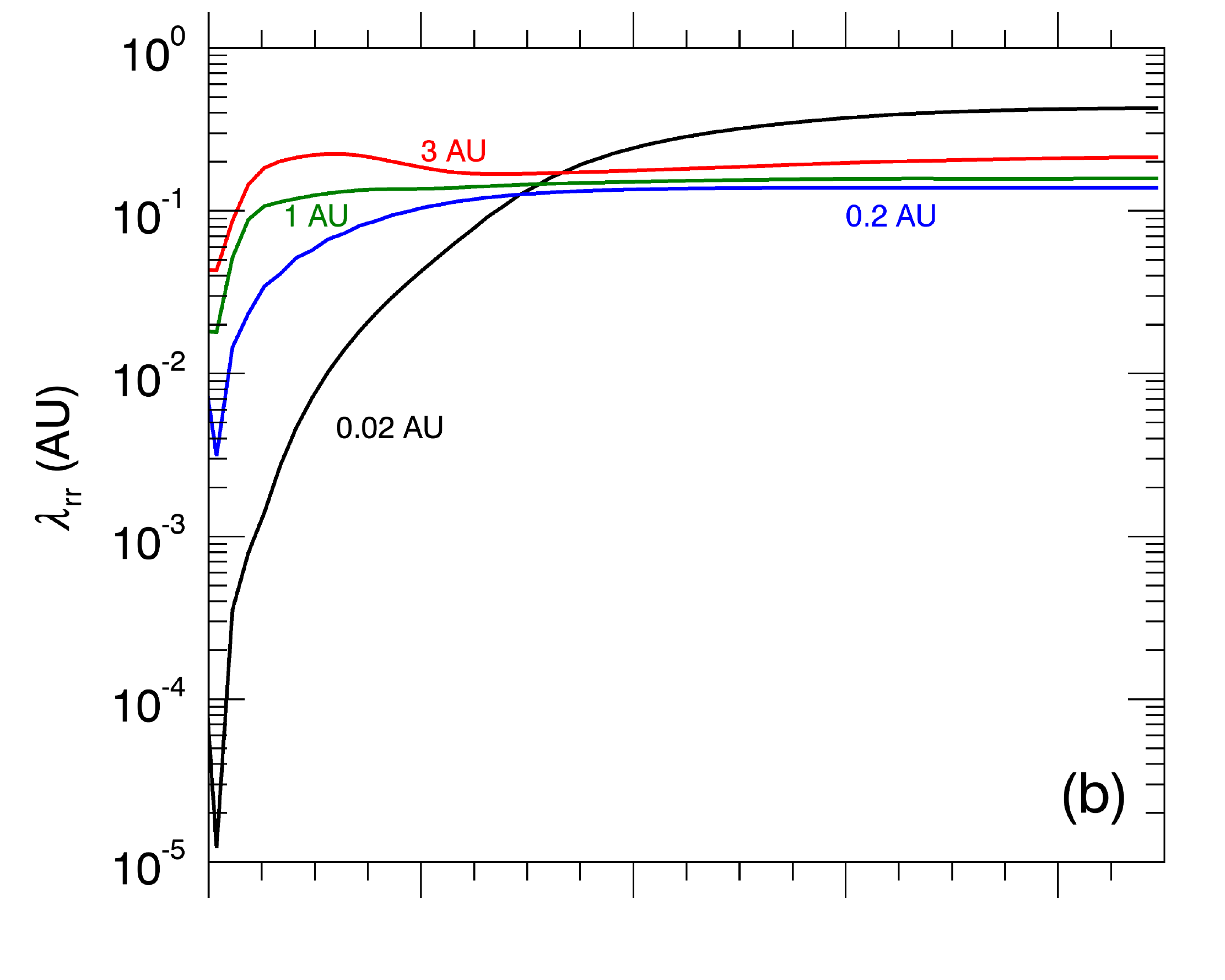}
\includegraphics[scale=.36]{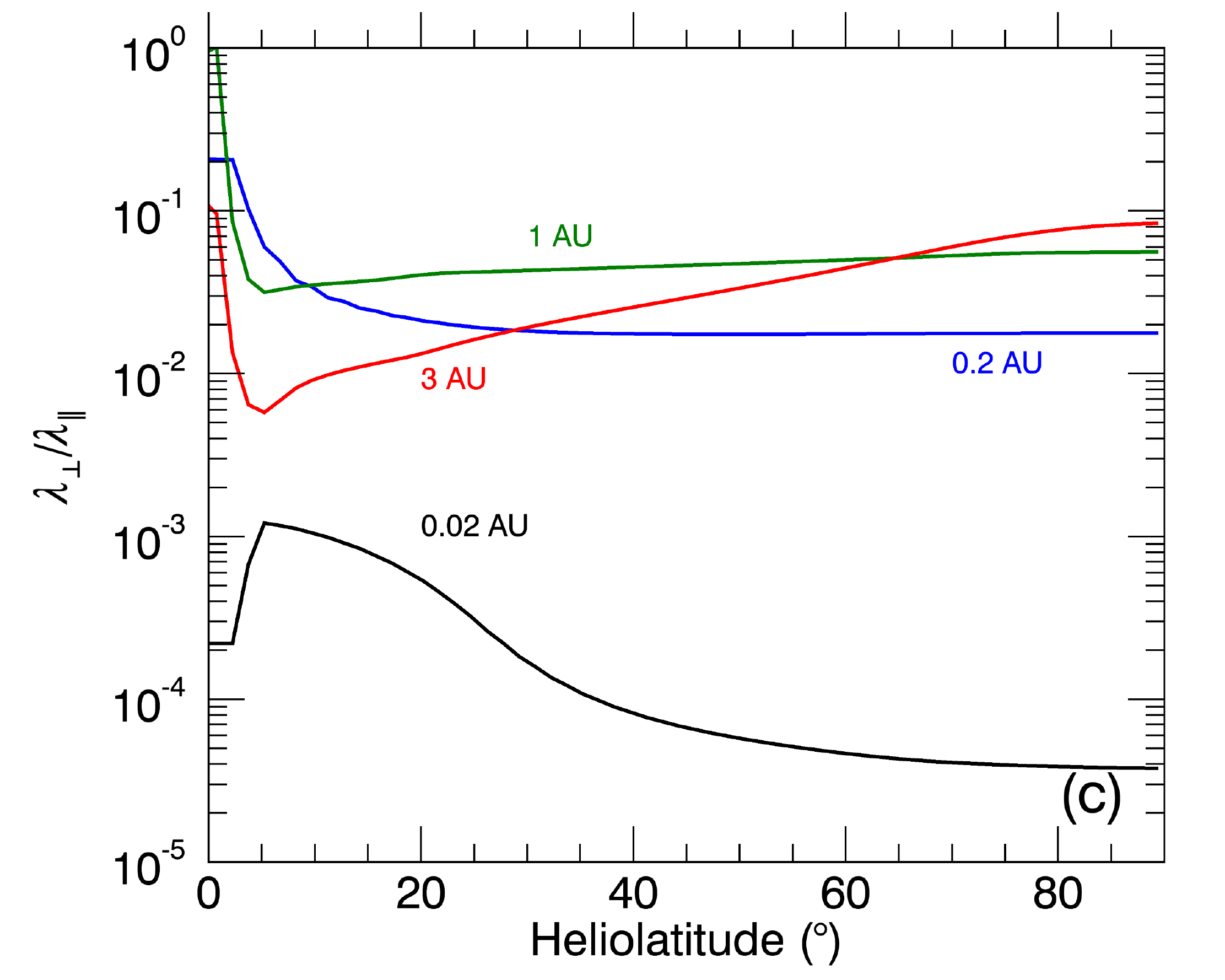}
\caption{The top panel (a) shows the latitudinal dependence of parallel (solid lines) and perpendicular (dashed lines) mfps. The middle (b) and bottom (c) panels show the latitudinal variation of $\lambda_{rr}$ and $\lambda_\perp / \lambda_\parallel$, respectively. All panels are for an untilted solar dipole and $p=2$. Black, blue, green, and red lines represent radial distances of 0.02, 0.2, 1, and 3 AU (4, 45, 215, and 645 $R_\odot$), respectively. Proton rigidity is 445 MV (100 MeV kinetic energy).}
\label{fig:mfp_lat}
\end{figure}

Figure \ref{fig:mfp_lat} shows the variation of mfps with latitude at different heliocentric distances for an untilted solar dipole. We see from Figure~\ref{fig:mfp_lat}a that, in general, $\lambda_\parallel$ (solid lines) increases by almost an order of magnitude as one leaves the solar equatorial plane and moves to higher latitudes, and assumes a near constant value as one approaches the polar regions. The opposite behaviour is seen for $\lambda_\perp$ (dashed lines), which decreases on moving away from the equatorial plane. This is a combined result of the increase in the IMF strength and the correlation scale of the turbulence ($\lambda$) while moving away from the solar equatorial plane (i.e., away from the HCS), and the increase in the turbulence energy due to shear-interactions between slow and fast solar winds. We note that very close to the sun (4 $R_\odot$, black line), $\lambda_\parallel$ first decreases with latitude as one leaves the equatorial plane, then increases at higher latitudes, to values larger even than those seen at larger heliocentric distances. This behavior is because of the IMF increasing monotonically with latitude, close to the sun. At larger distances, the IMF plateaus with increasing latitude, and from 1 AU onwards it decreases in the polar regions (See Figure \ref{fig:ingred2}). Thus, particles experience less scattering in polar regions close to the sun. This also explains the latitudinal variation of $\lambda_\perp$ at 4 $R_\odot$. 

Figure \ref{fig:mfp_lat}b shows the increase in $\lambda_{rr}$ as one moves towards the polar regions, and illustrates once again the fact that while $\lambda_{rr}$ is affected by $\lambda_\perp$ very close to the sun at low latitudes, near the polar regions it follows the trend set by $\lambda_\parallel$. Figure \ref{fig:mfp_lat}c shows that the ratio of $\lambda_\perp$ to $\lambda_\parallel$ decreases as one leaves the solar equatorial plane (i.e., away from the HCS), with the perpendicular mfp staying 1-2 orders of magnitude below the parallel mfp, except very close to the sun (4 $R_\odot$, black line) where it becomes 3 orders of magnitude smaller because of the low turbulence levels in that region. We will examine the latitudinal dependence of the mfps once again in meridional plane figures in Section 4.5, below.

\subsection{Rigidity dependence of mfps}

\begin{figure}
\includegraphics[scale=.36]{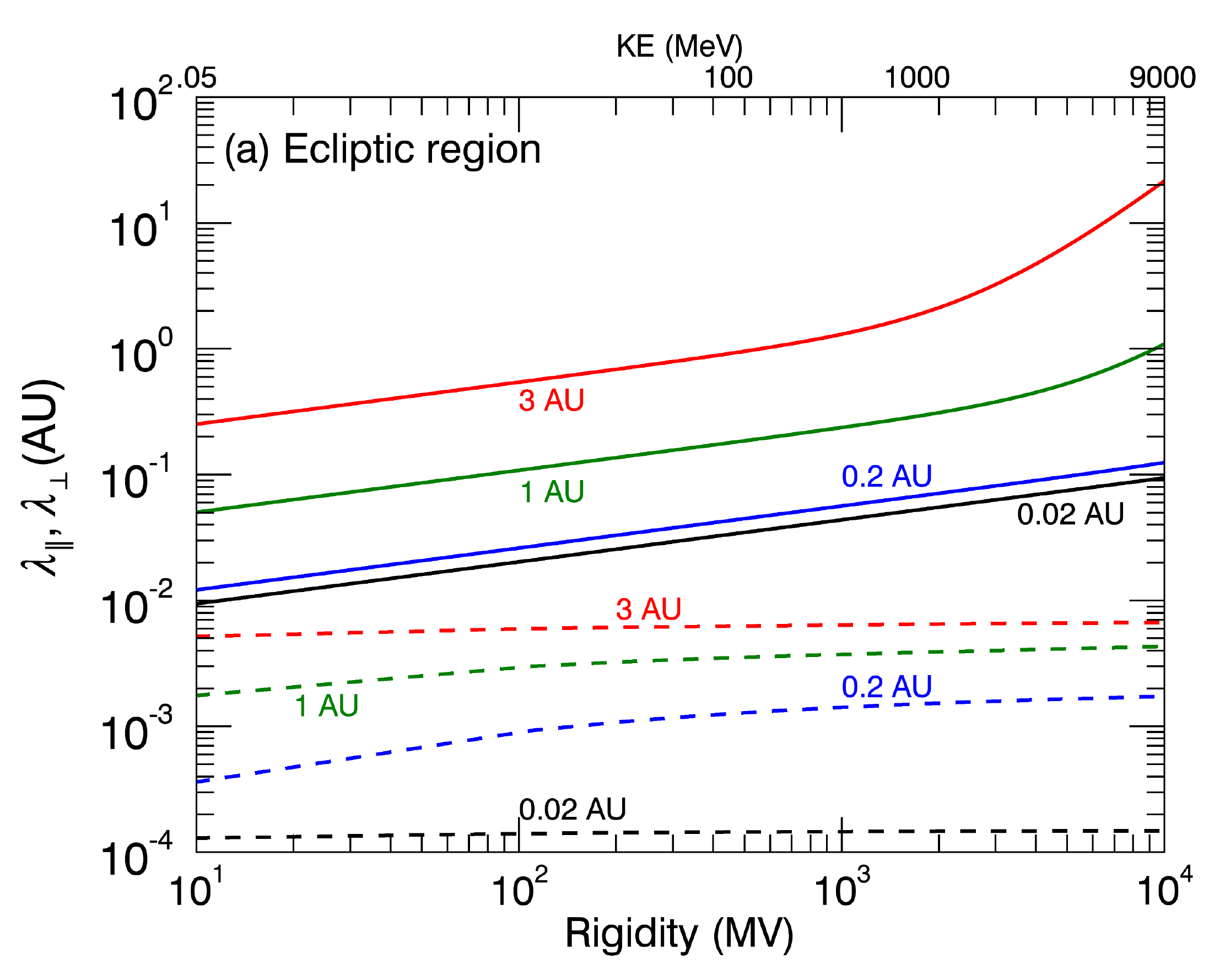}
\includegraphics[scale=.36]{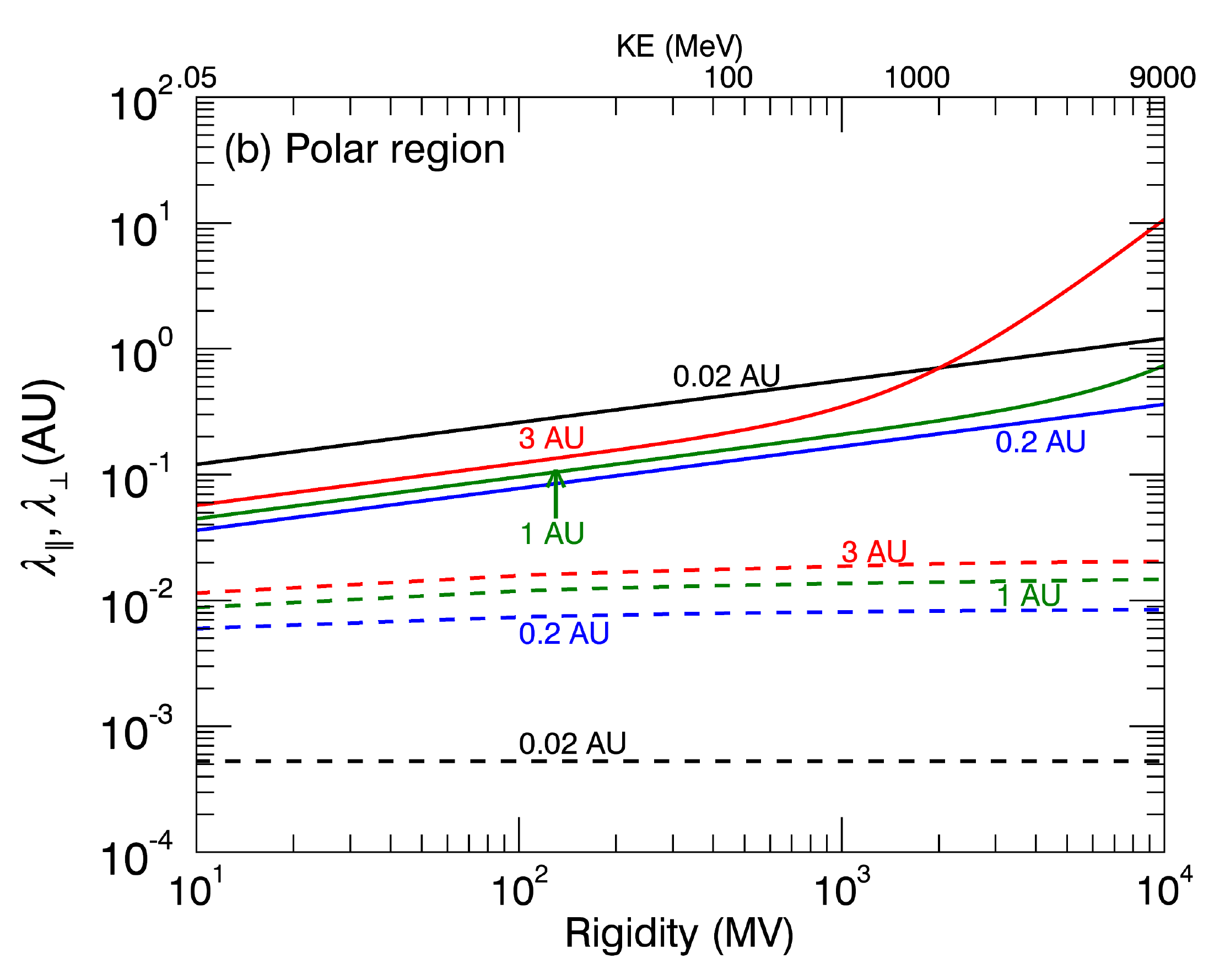}
\caption{Rigidity dependence of $\lambda_\parallel$ (solid line) and $\lambda_\perp$ (dashed line), (a) near the ecliptic plane ($7\degree$ heliolatitude), and (b) in the polar regions ($86\degree$ heliolatitude), for an untilted solar dipole and $p=2$. Black, blue, green, and red lines represent radial distances of 0.02, 0.2, 1, and 3 AU (4, 45, 215, and 645 $R_\odot$), respectively.}
\label{fig:rig}
\end{figure}

In Figure \ref{fig:rig} we plot the rigidity ($P$) dependence of mfps for protons at different radial distances in the ecliptic and polar regions. Below 1 AU, $\lambda_\parallel \propto P^{0.33}$ for all rigidities considered here ($10 - 10^4$ MV). Above 1 AU there is a steepening of the slope for rigidities larger than $10^3$ MV. As noted in Section 2.1, this is due to high energy particles resonating with turbulent fluctuations in the energy containing range instead of the inertial range. As the IMF ($B$) decreases with heliocentric distance, a high rigidity particle's Larmor radius ($R_L = P/Bc$) may become resonant with the correlation scale of the turbulence ($\lambda_s$). When $R_L/\lambda_s >> 1$, the expression in braces in Equation~\eqref{eq:mfp_p} scales with rigidity as $P^{5/3}$, and we have $\lambda_\parallel \propto P^2$ instead of $\lambda_\parallel \propto P^{1/3}$. Indeed, for rigidities $\sim 10^4$ MV we find that $\lambda_\parallel \propto P^{1.2}$ at 1 AU and $\lambda_\parallel \propto P^{1.8}$ at 3 AU \citep[See also the discussion on the effect of pickup ion driven turbulence on high-rigidity particles in the outer heliosphere in][]{zank1998radial}. Our results agree well with the observations shown in \cite{bieber1994proton}, with power indices ranging from 0.2 to 0.56 for a number of solar events where rigidity ranges from 10 to $10^3$ MV. Our results also agree with the theoretical and numerical findings in \cite{bieber1994proton} and \cite{pei2010cosmic}. 

In general, $\lambda_\perp$ shows lower variation with rigidity. In the polar regions $\lambda_\perp$ stays nearly constant with rigidity. This behavior is consistent with the finding of \cite{bieber2004GRL} that NLGC predicts a very weak rigidity dependence, and they note that this is supported by observations for rigidities between $10^2 - 10^4$ MV. Note that the rigidity profiles of $\lambda_\parallel$ and $\lambda_\perp$ that we derive from simulation results and diffusion theories are quite different from some that have been employed in the literature to model solar modulation of Galactic cosmic rays \citep[e.g., see Figure 12 of][]{vos2015ApJ815}.

\subsection{Meridional plane contours}

\begin{figure}
\includegraphics[scale=.445]{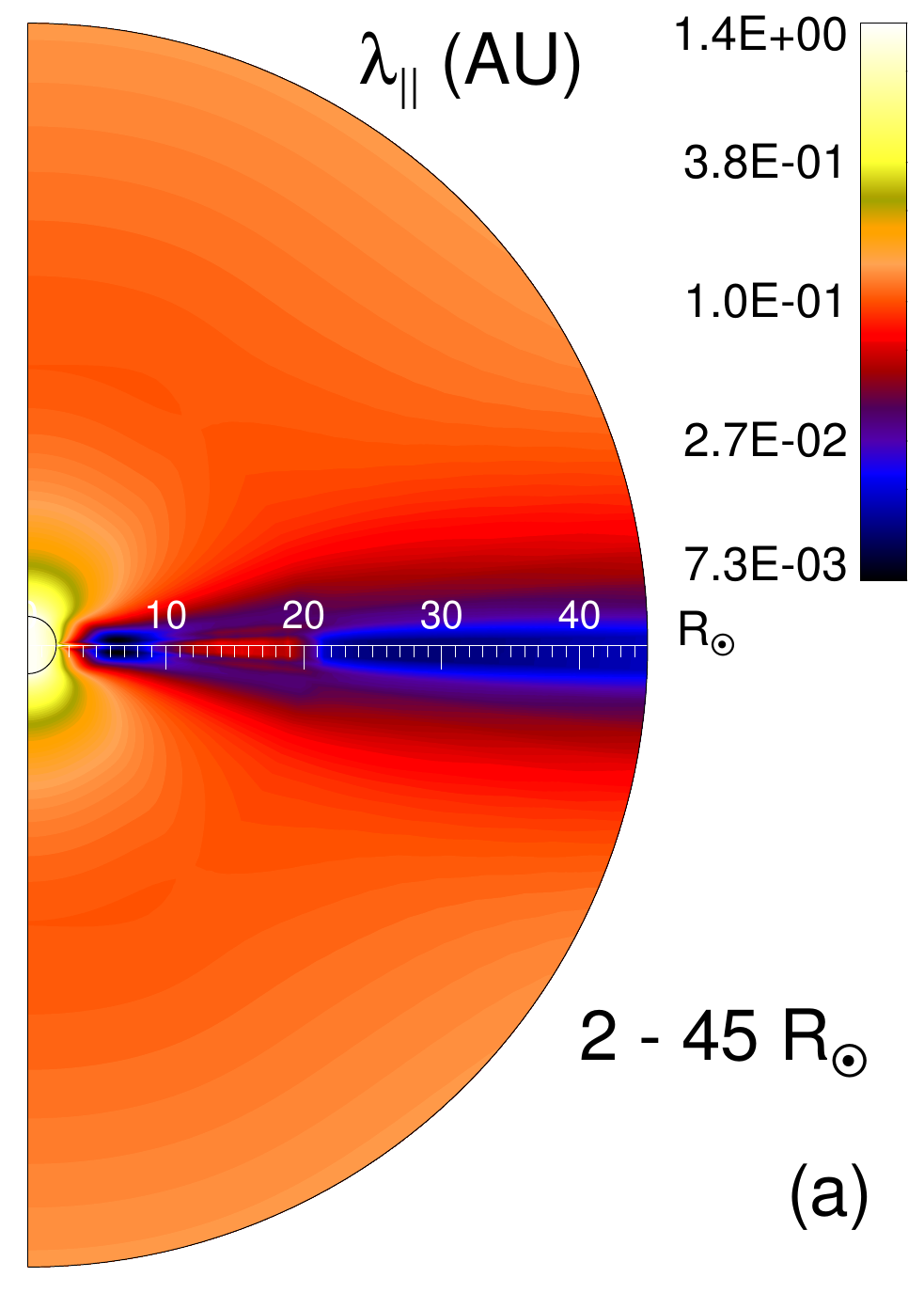}
\includegraphics[scale=.445]{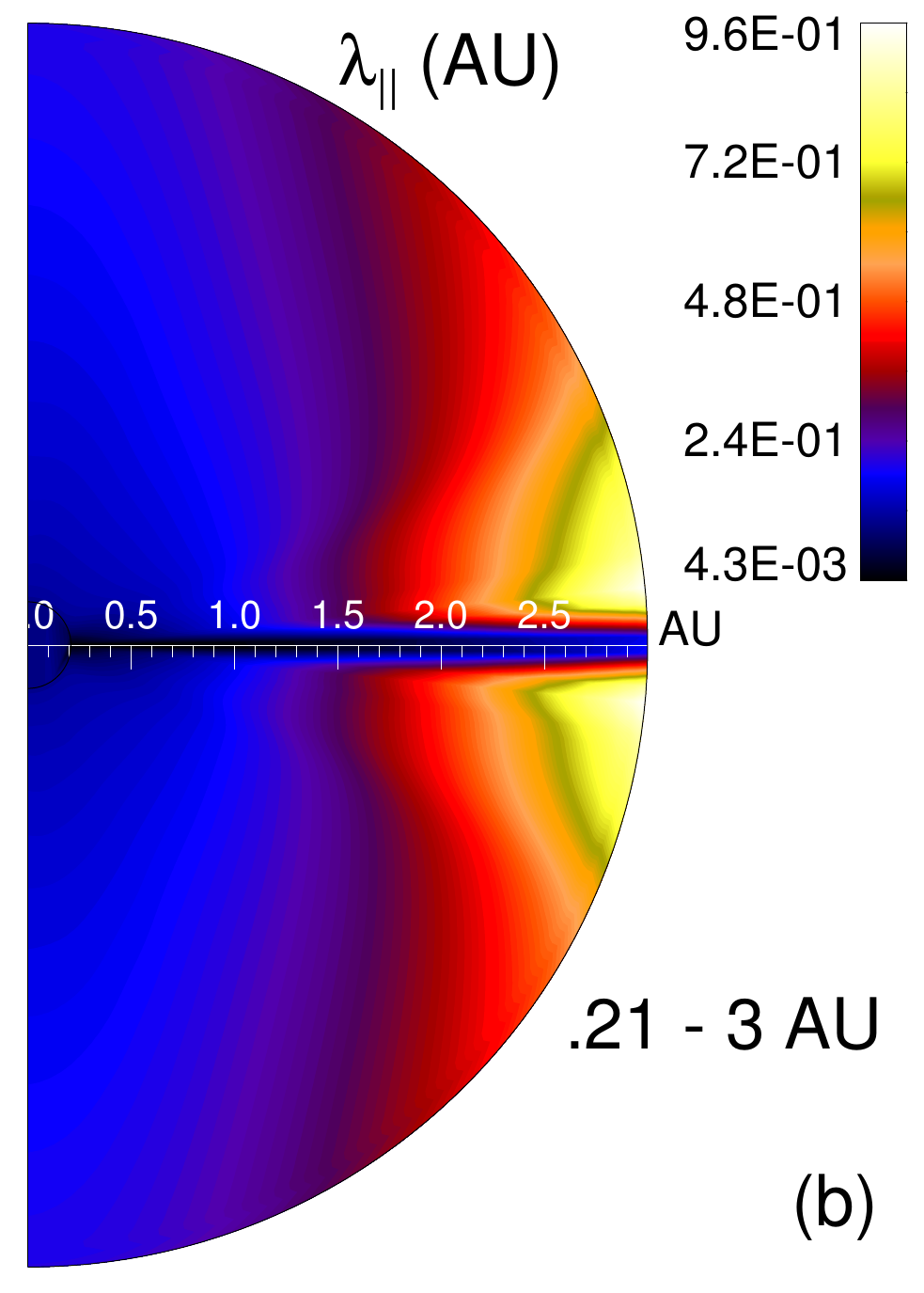}
\includegraphics[scale=.445]{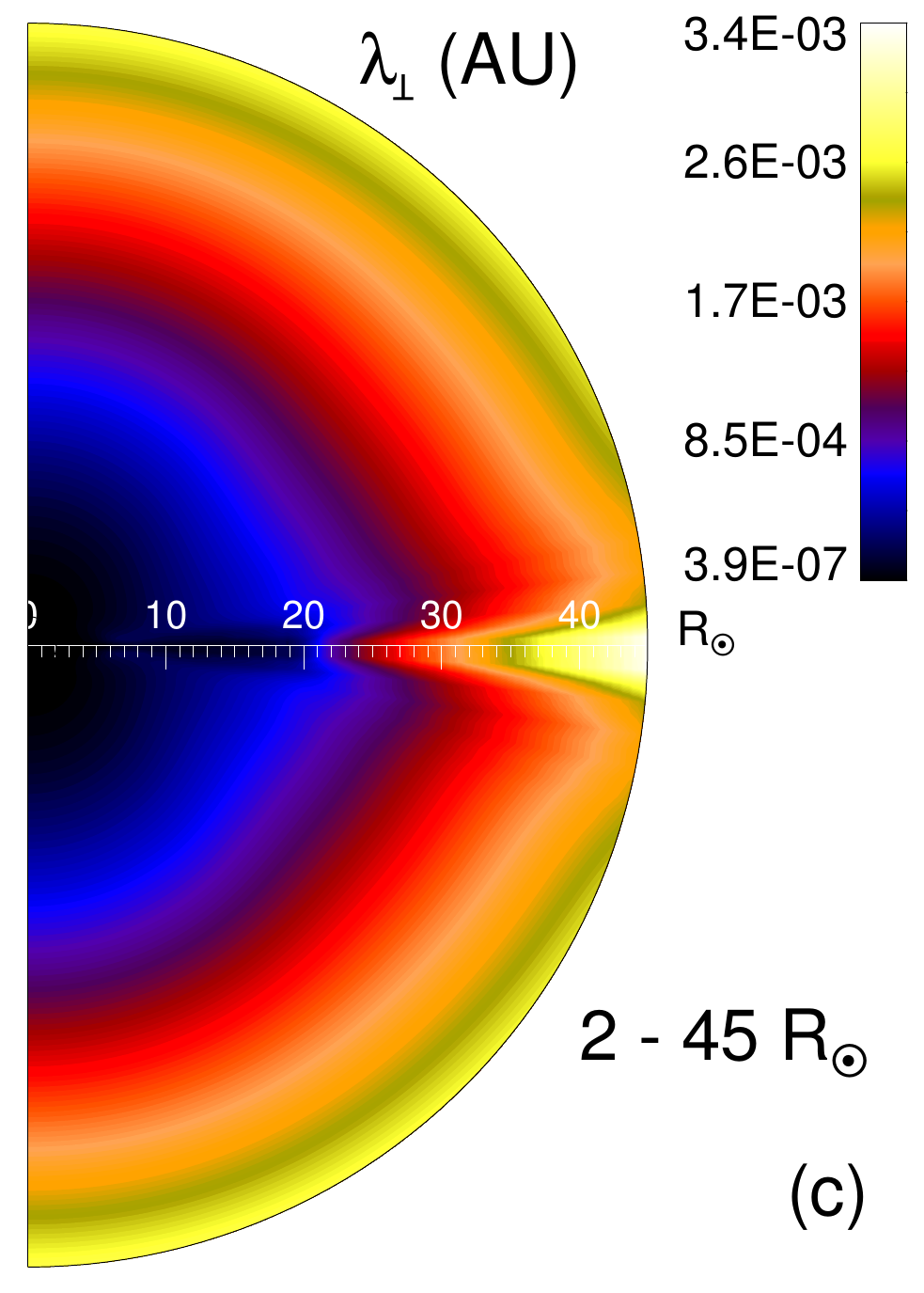}
\includegraphics[scale=.445]{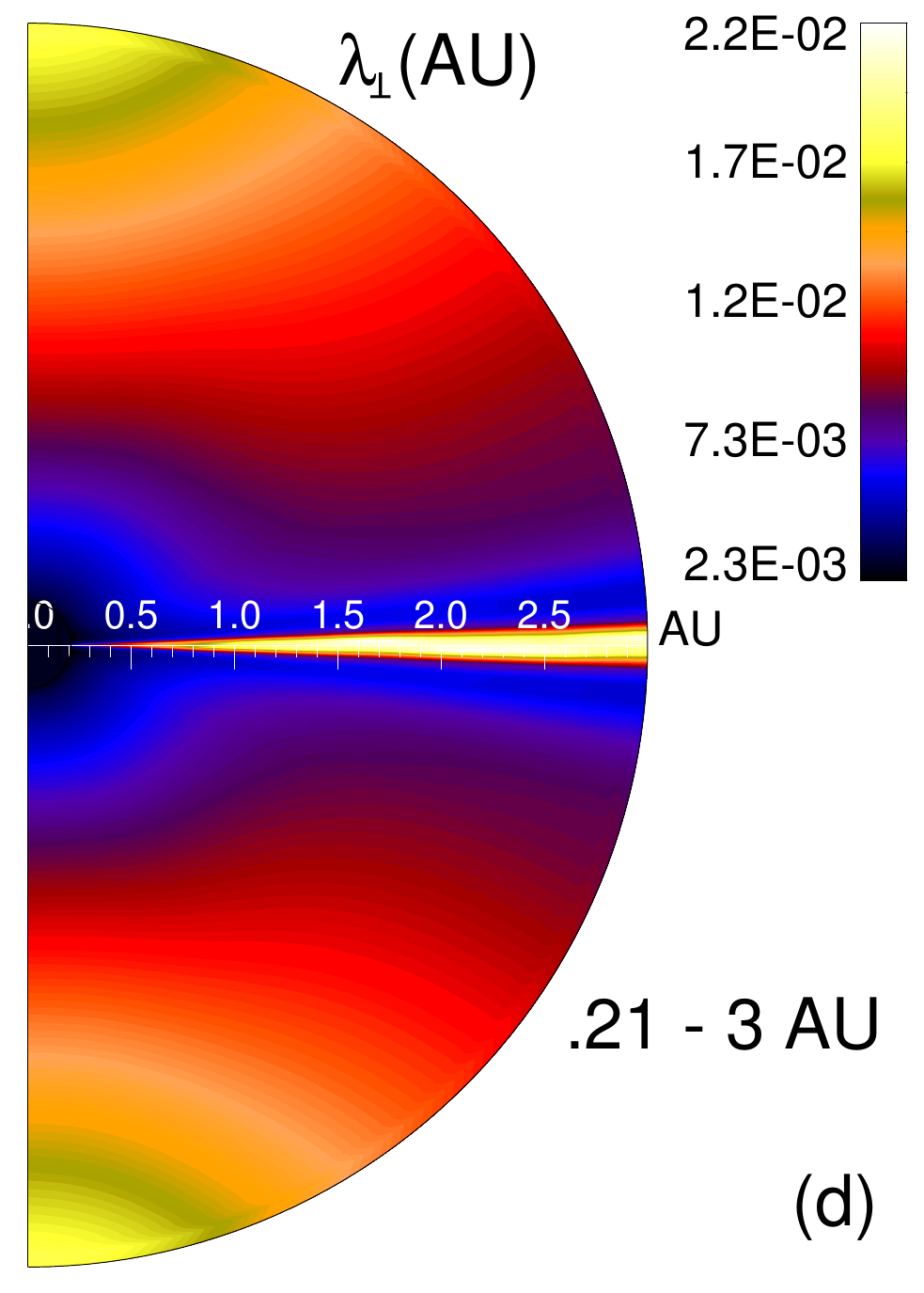}
\includegraphics[scale=.445]{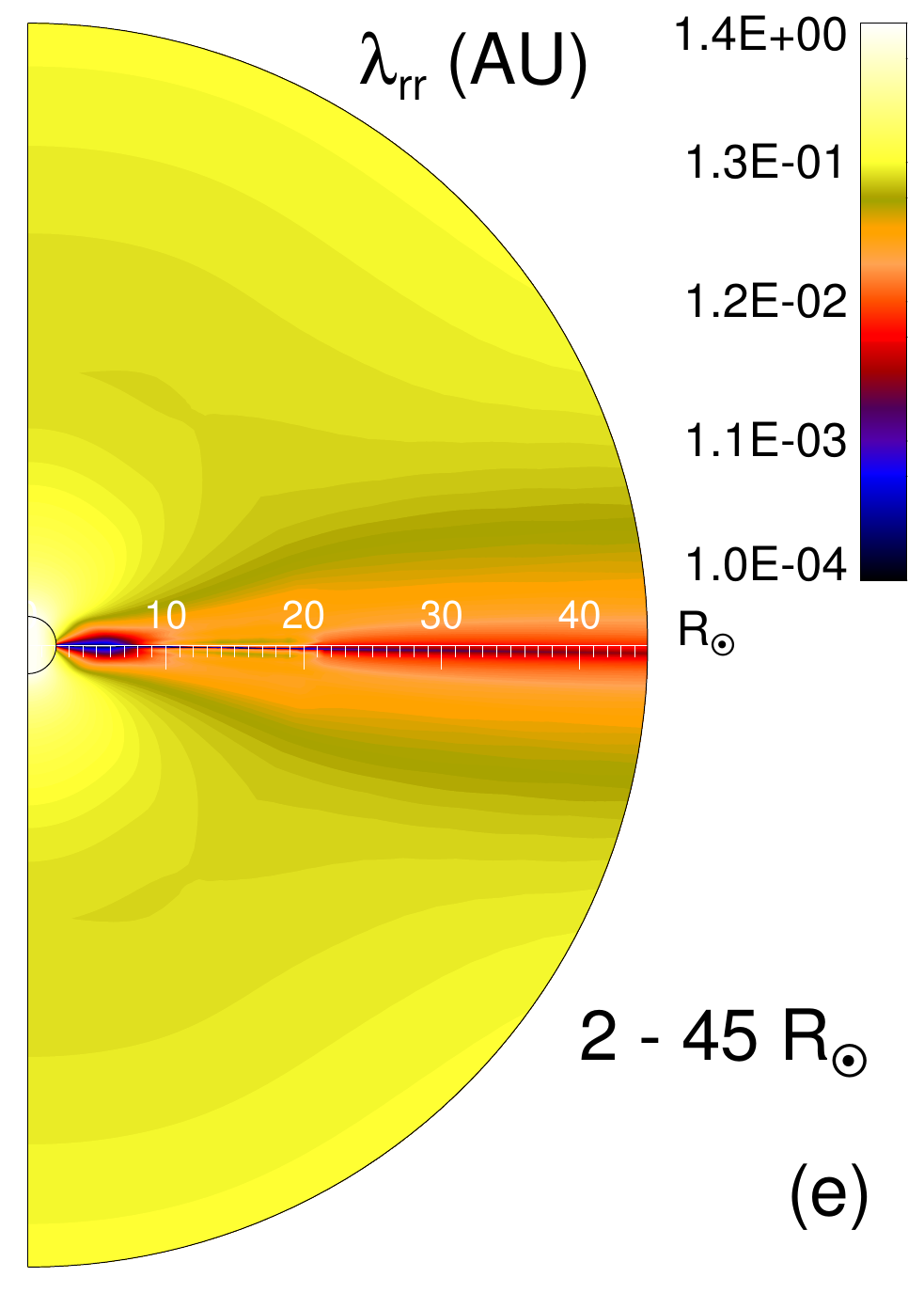}
\includegraphics[scale=.445]{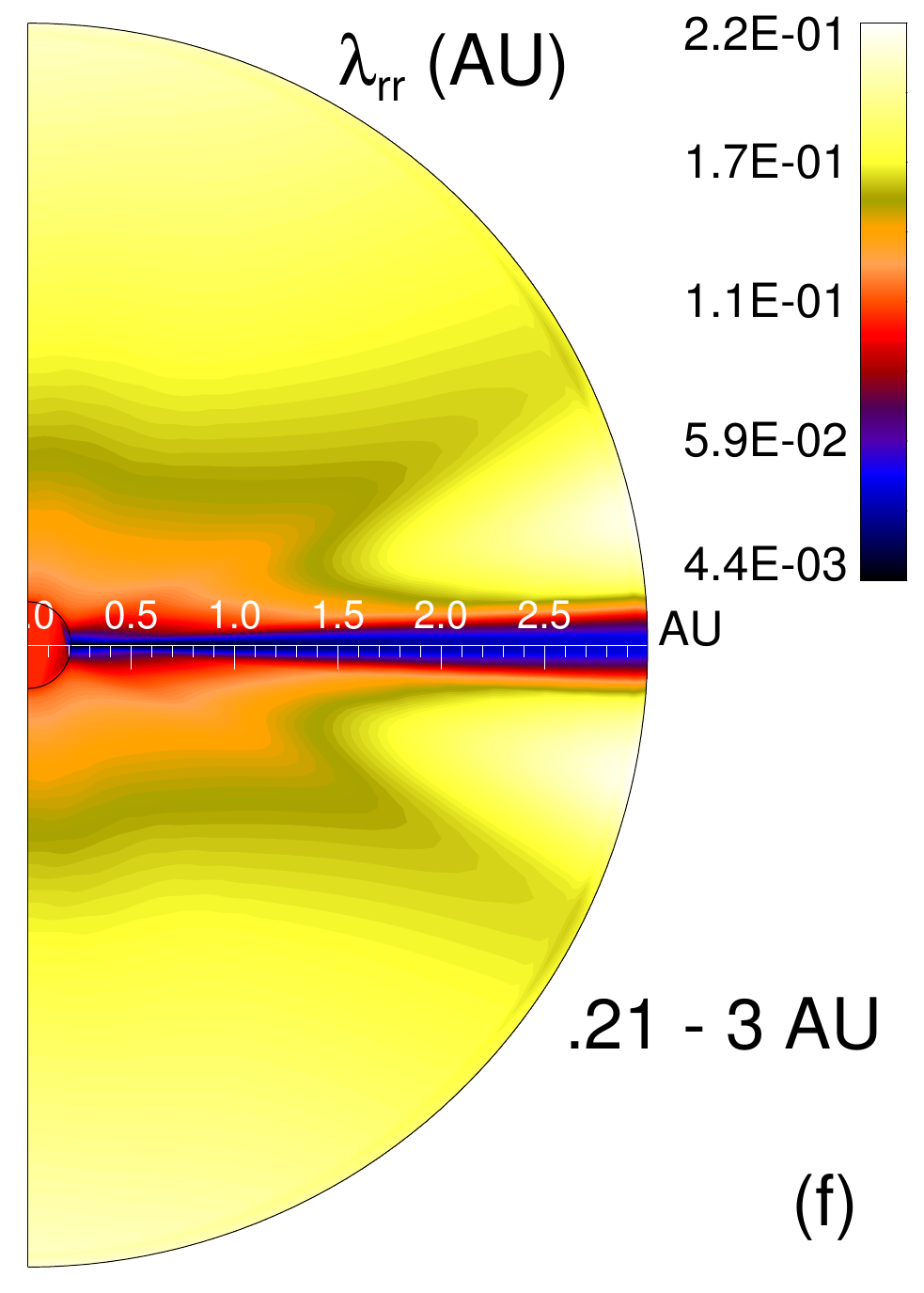}
\includegraphics[scale=.445]{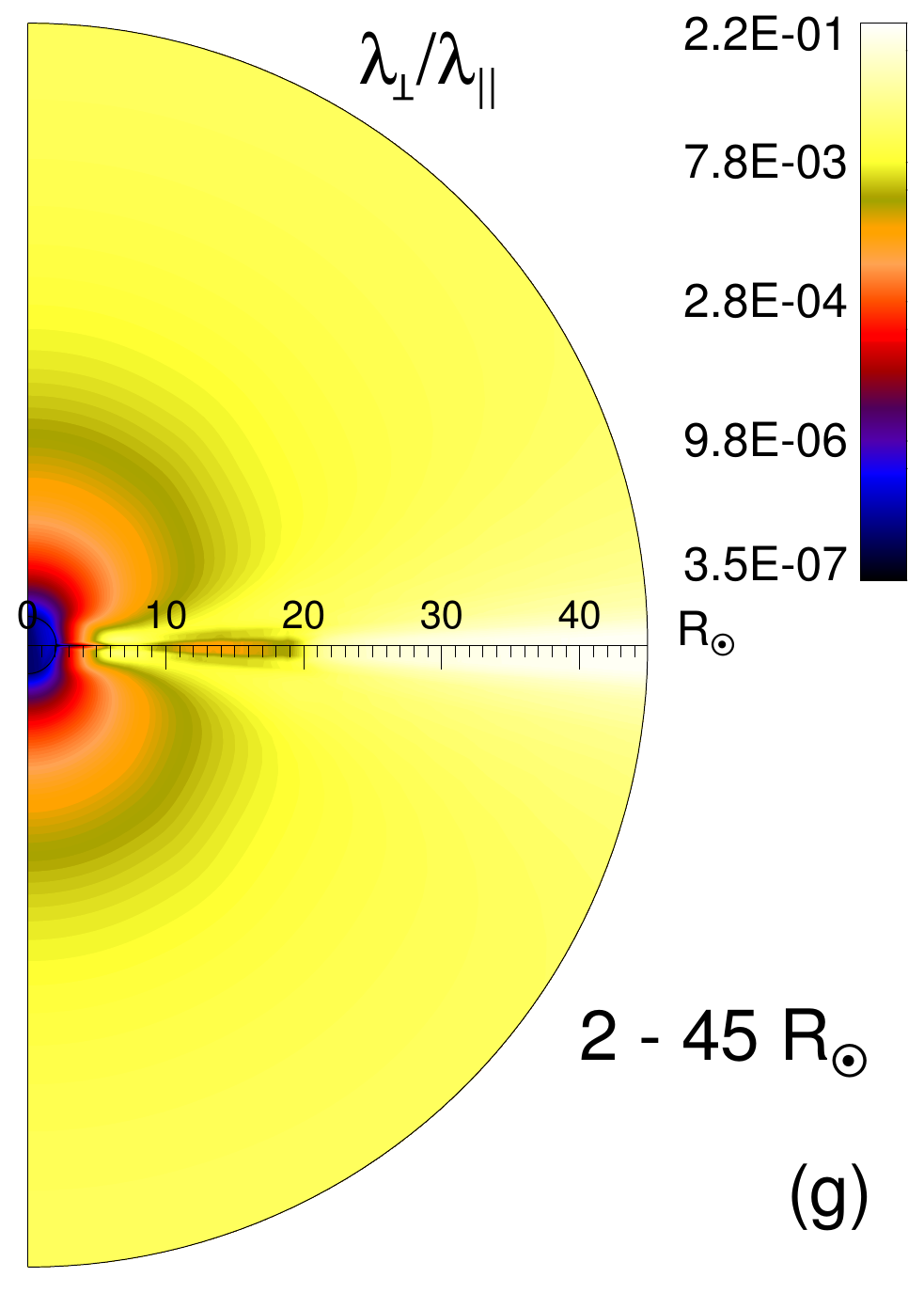}
\includegraphics[scale=.445]{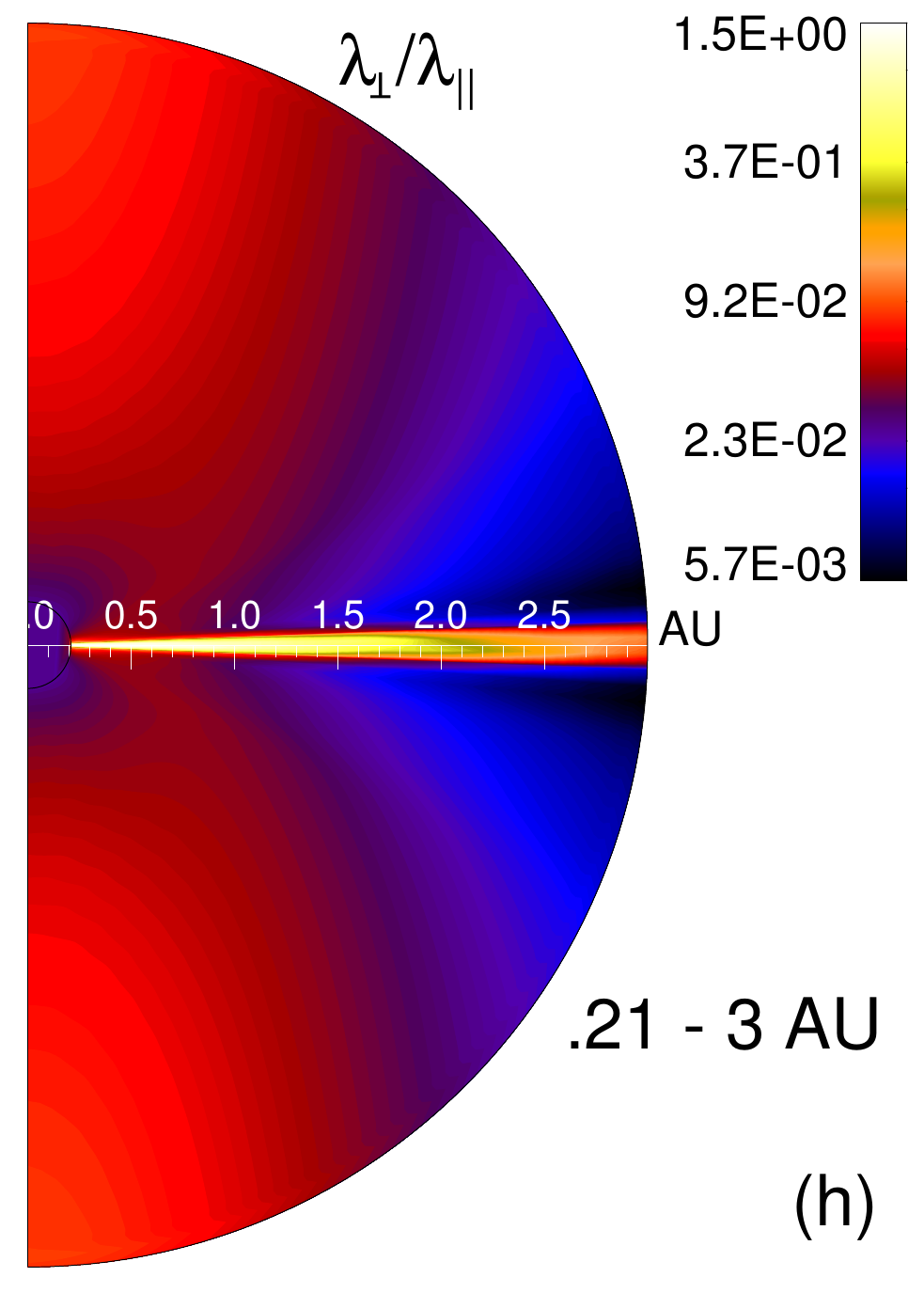}
\caption{Contour plots in the meridional plane of mfps, with a solar dipole that is untilted with respect to the solar rotation axis. The inner and intermediate regions ($2 - 45~R_\odot$) and the outer region ($0.21 - 3$ AU, or $45 - 645~R_\odot$) are shown separately. Proton rigidity is 445 MV (100 MeV kinetic energy) and $p=2$.}
\label{fig:merid_untilt}
\end{figure}

In this section, we describe the variation of $\lambda_\parallel, \lambda_\perp, \lambda_{rr}$, and $\lambda_\perp/\lambda_\parallel$ in meridional planes for 100 MeV protons, complementing results of the previous sections. Figure \ref{fig:merid_untilt} shows results from a simulation with a source magnetic dipole that is untilted with respect to the solar rotation axis. It is clear that at the HCS, with its vanishing magnetic field, perpendicular diffusion is comparable to parallel diffusion in most of the inner heliosphere, with $\lambda_\perp$ and $\lambda_\parallel$ both around  $0.01$ AU. In the broader ecliptic plane, however, $\lambda_\parallel$ remains 1-2 orders of magnitude above $\lambda_\perp$, varying from 0.01 to almost 1 AU within a radial distance of 10 $R_\odot$ to 3 AU, while $\lambda_\perp$ increases from $\sim 0.0001$ to $0.01$ AU. As noted in the 1-D plots, very close to the sun $\lambda_\parallel$ experiences a dramatic increase to a value of 1 AU due to the weak turbulence and strong magnetic field prevailing there. 

We also see that at radial distances of $1.5 - 3$ AU, $\lambda_\parallel$ is a few times larger at lower latitudes, compared to values in polar regions. This is because the IMF decreases and the turbulence energy increases with latitude at these radial distances, leading to a reduction in parallel diffusion in the polar regions, and a corresponding increase in perpendicular diffusion. This can also be seen in Figure~\ref{fig:merid_untilt}h showing contours of $\lambda_\perp/\lambda_\parallel$, which increases by nearly one order of magnitude from low latitudes to the poles. The radial mfp increases uniformly with heliocentric distance at lower latitudes, but is dominated by $\lambda_\parallel$ in polar regions, because of the small winding angle between the IMF and the radial direction here. This leads to $\lambda_{rr}$ acquiring a nearly constant value of around 0.2 AU in polar regions beyond 2 AU. 

\begin{figure}
\includegraphics[scale=.445]{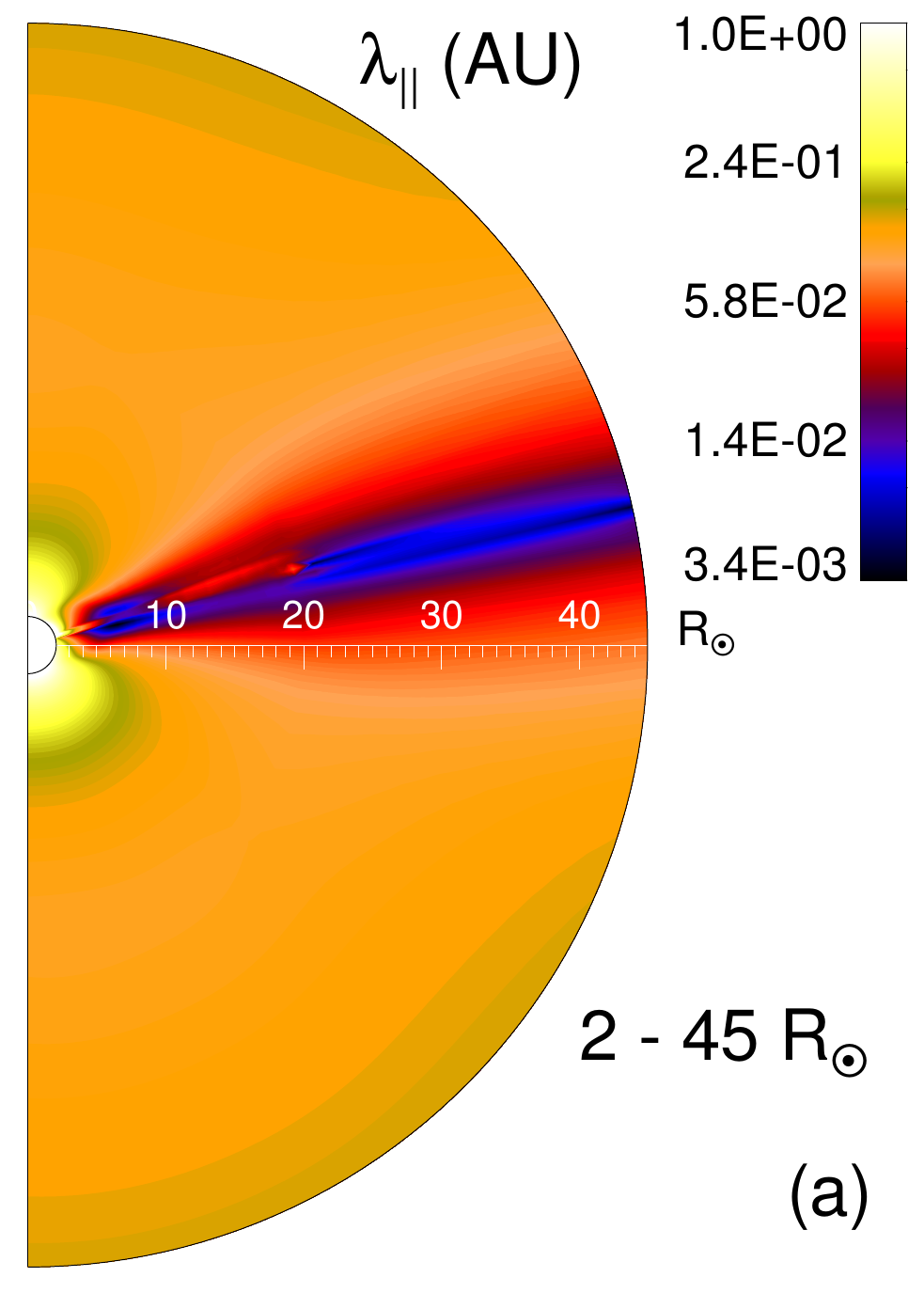}
\includegraphics[scale=.445]{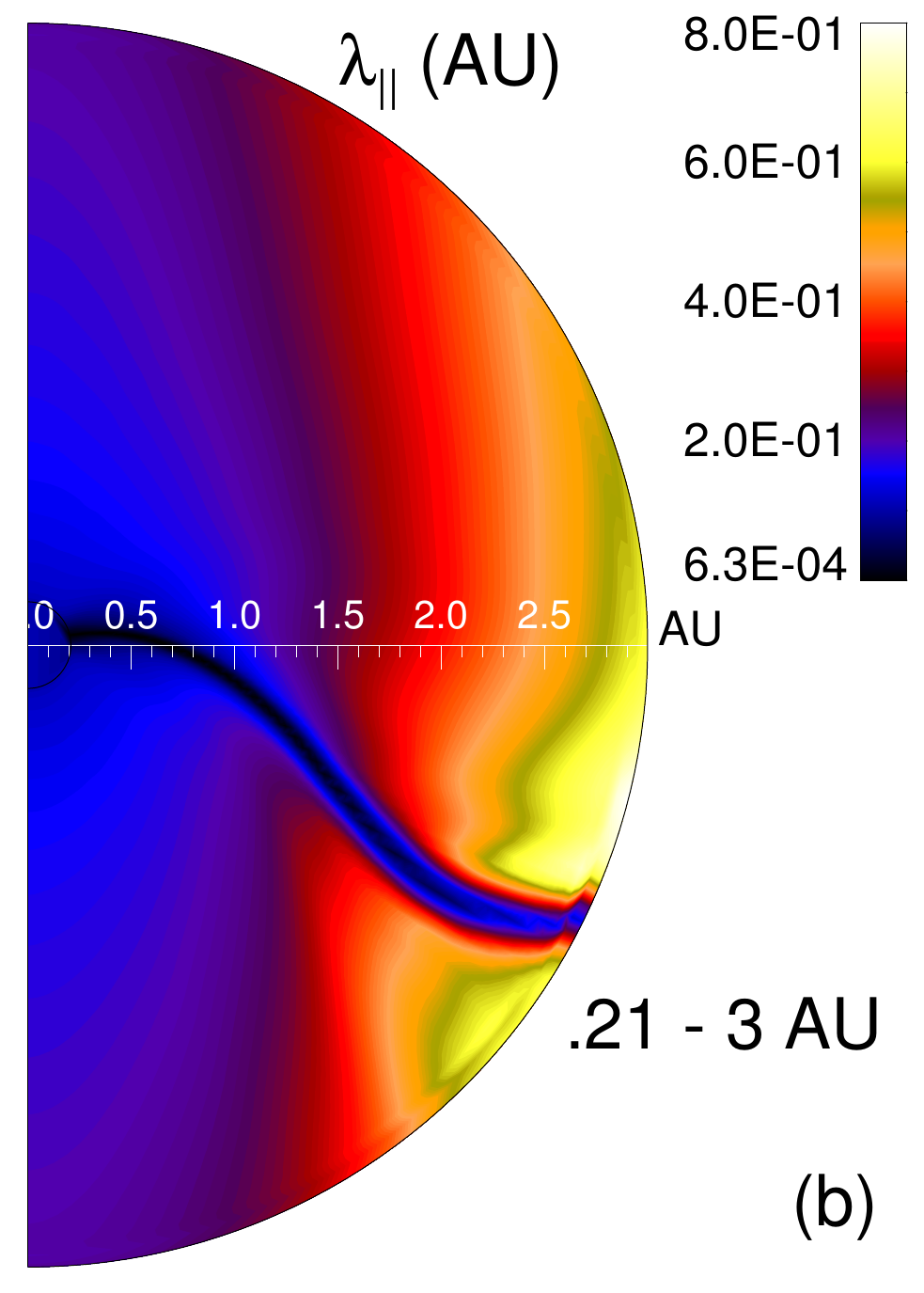}
\includegraphics[scale=.445]{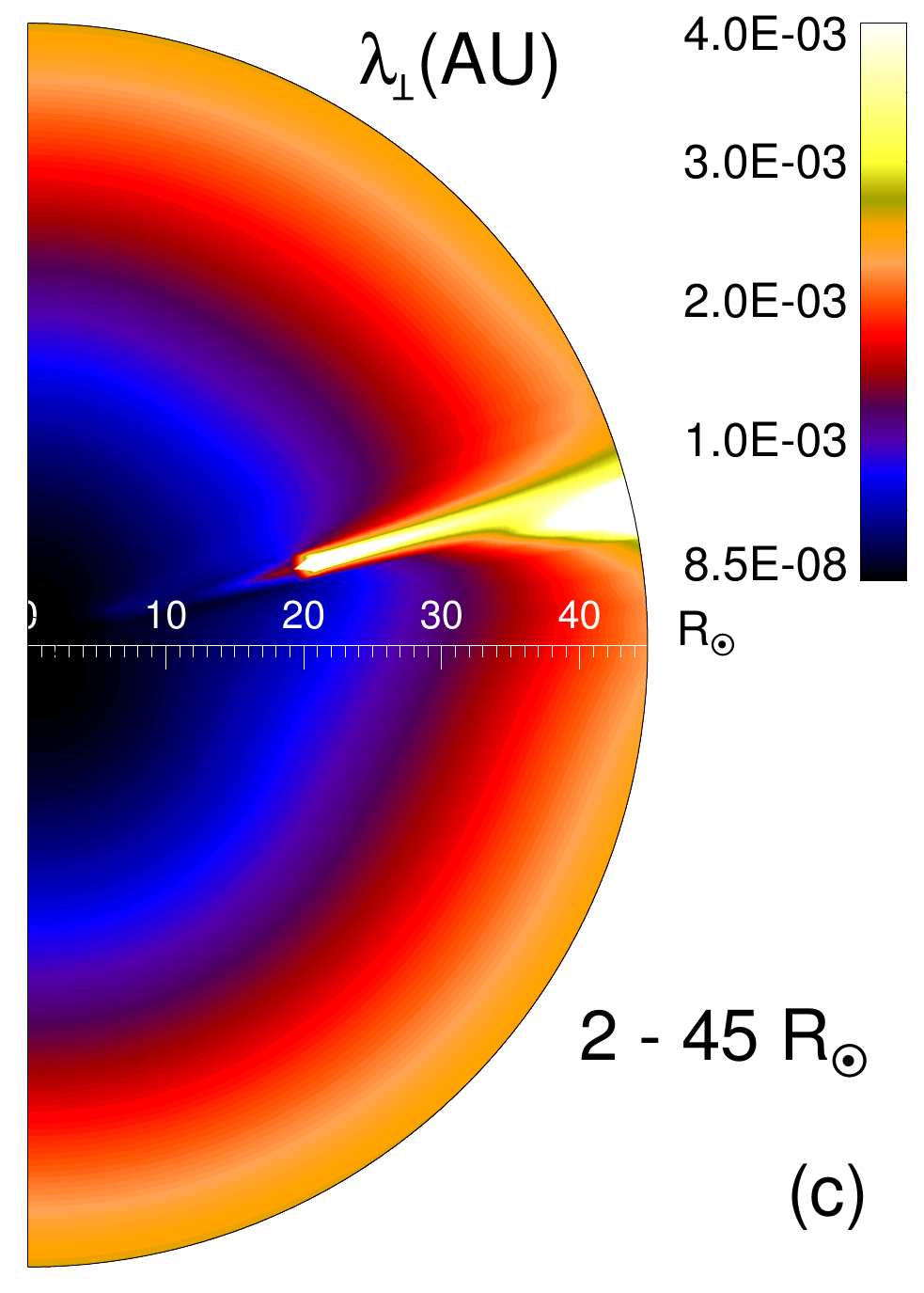}
\includegraphics[scale=.445]{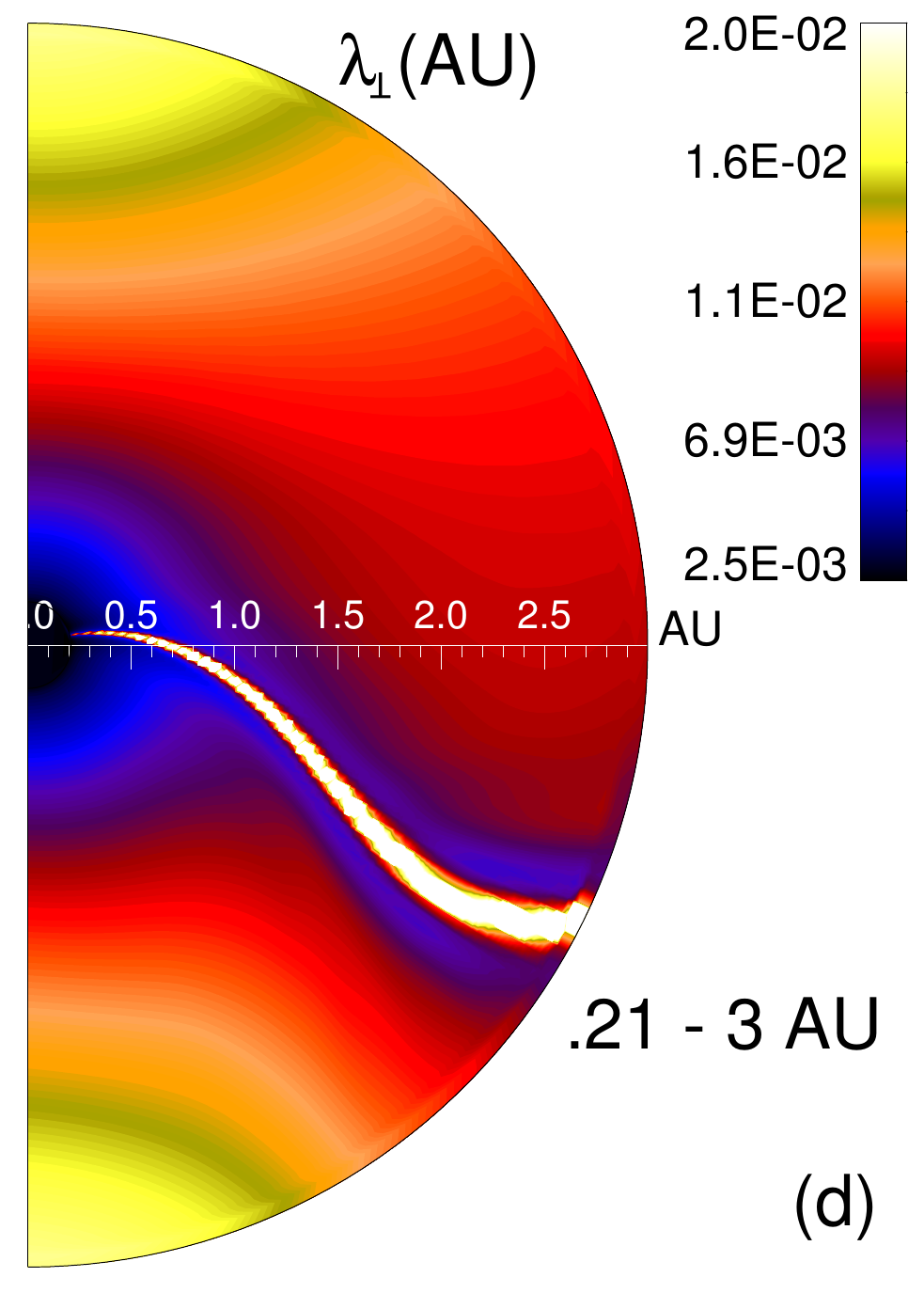}
\includegraphics[scale=.445]{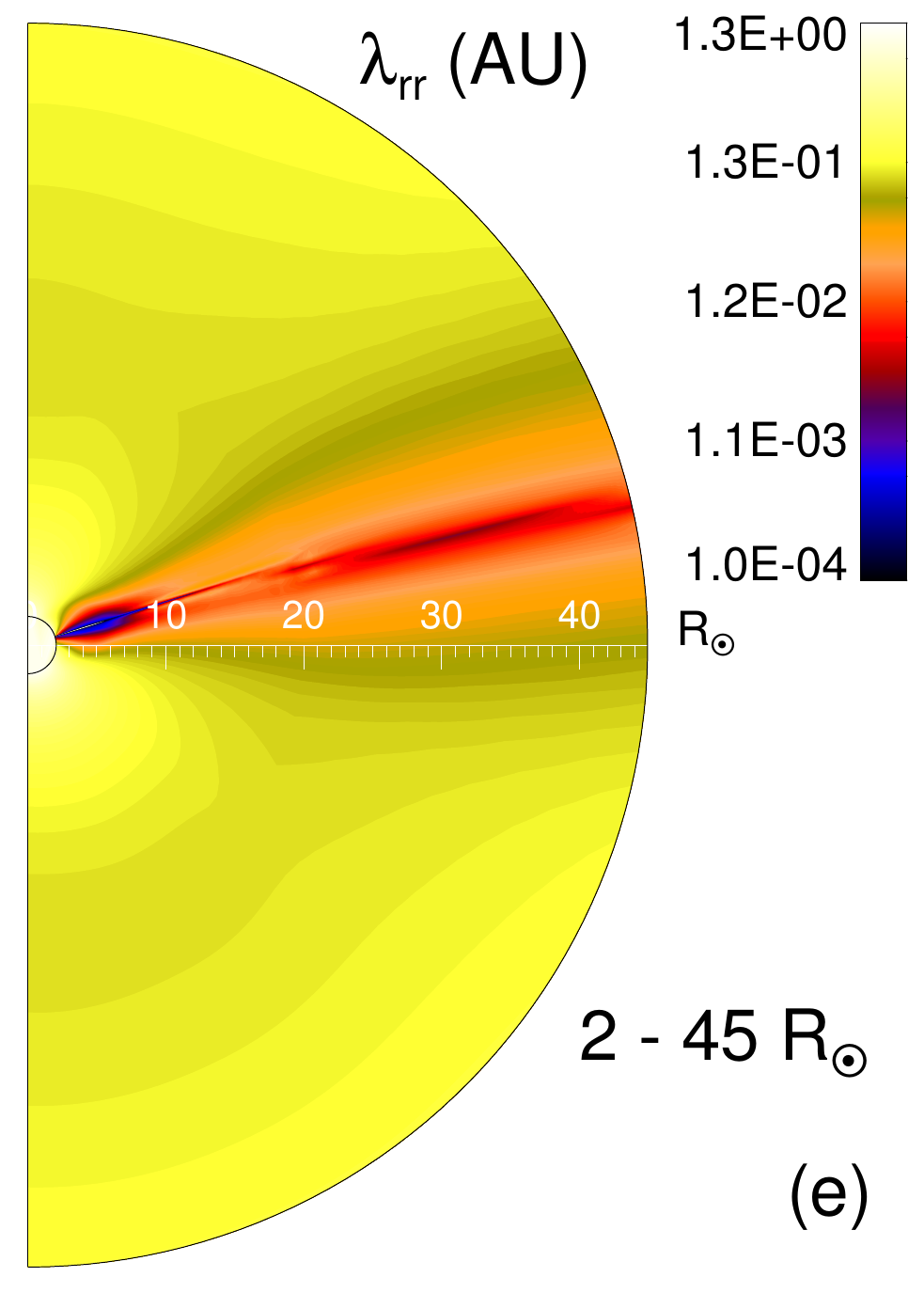}
\includegraphics[scale=.445]{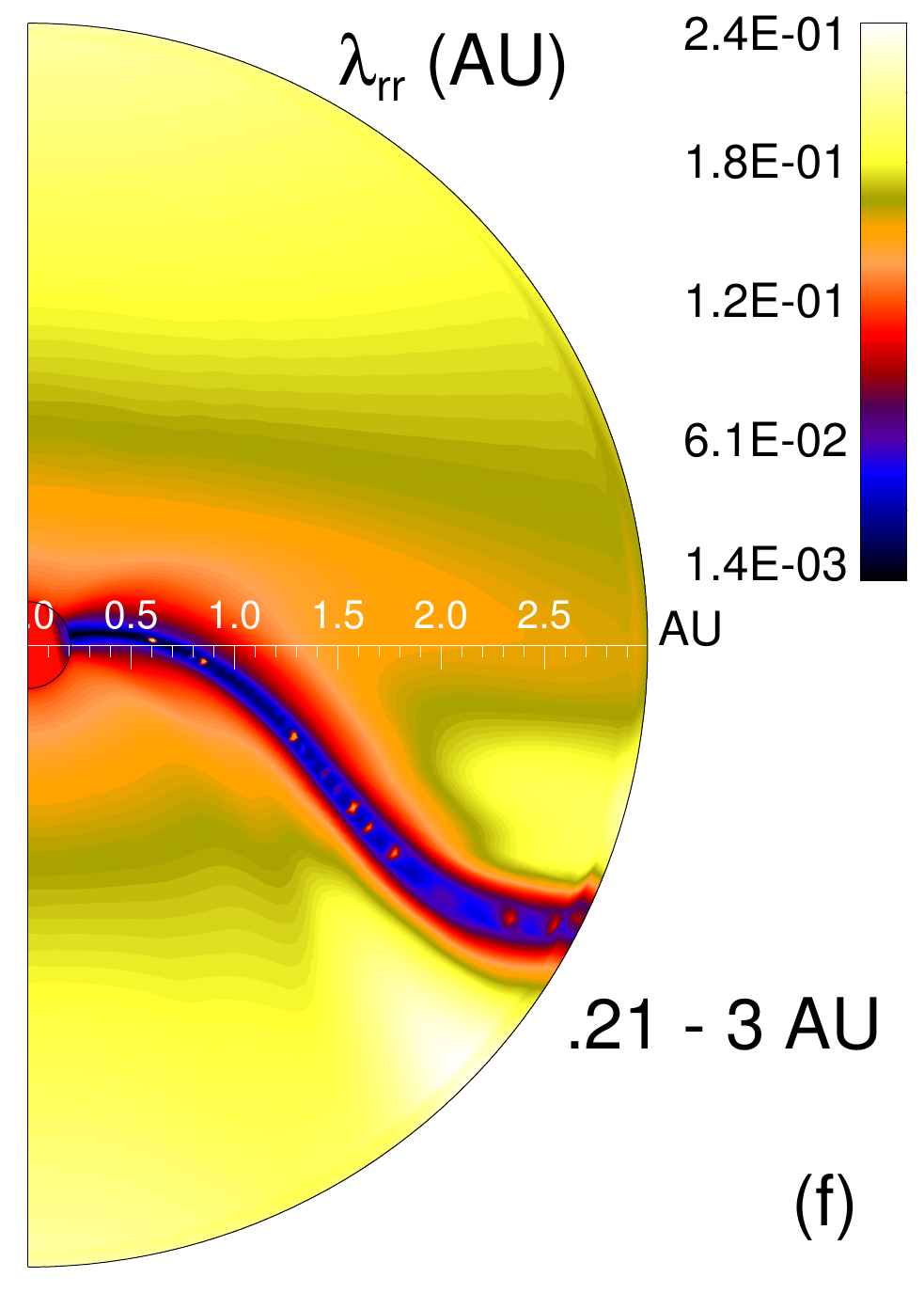}
\includegraphics[scale=.445]{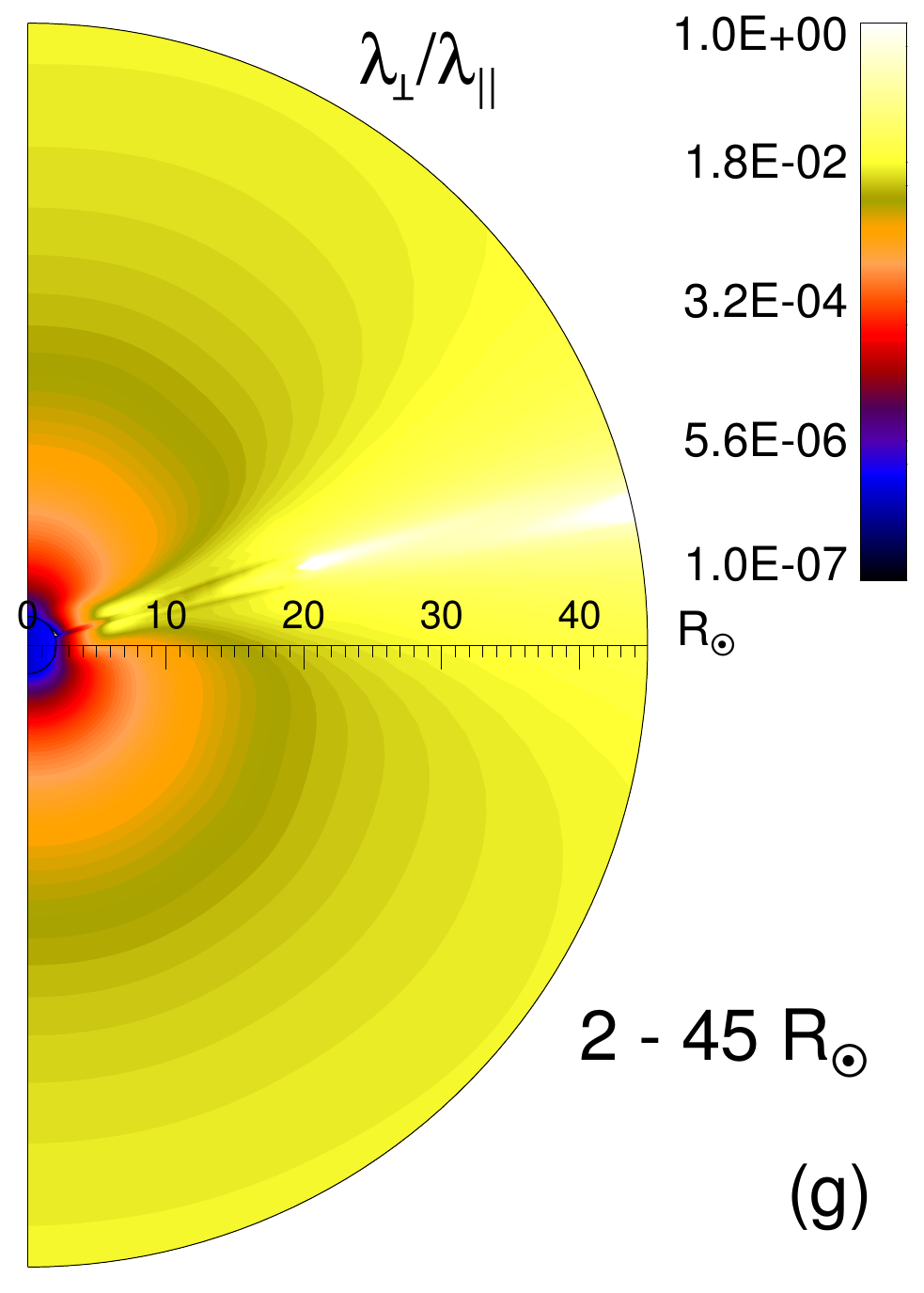}
\includegraphics[scale=.445]{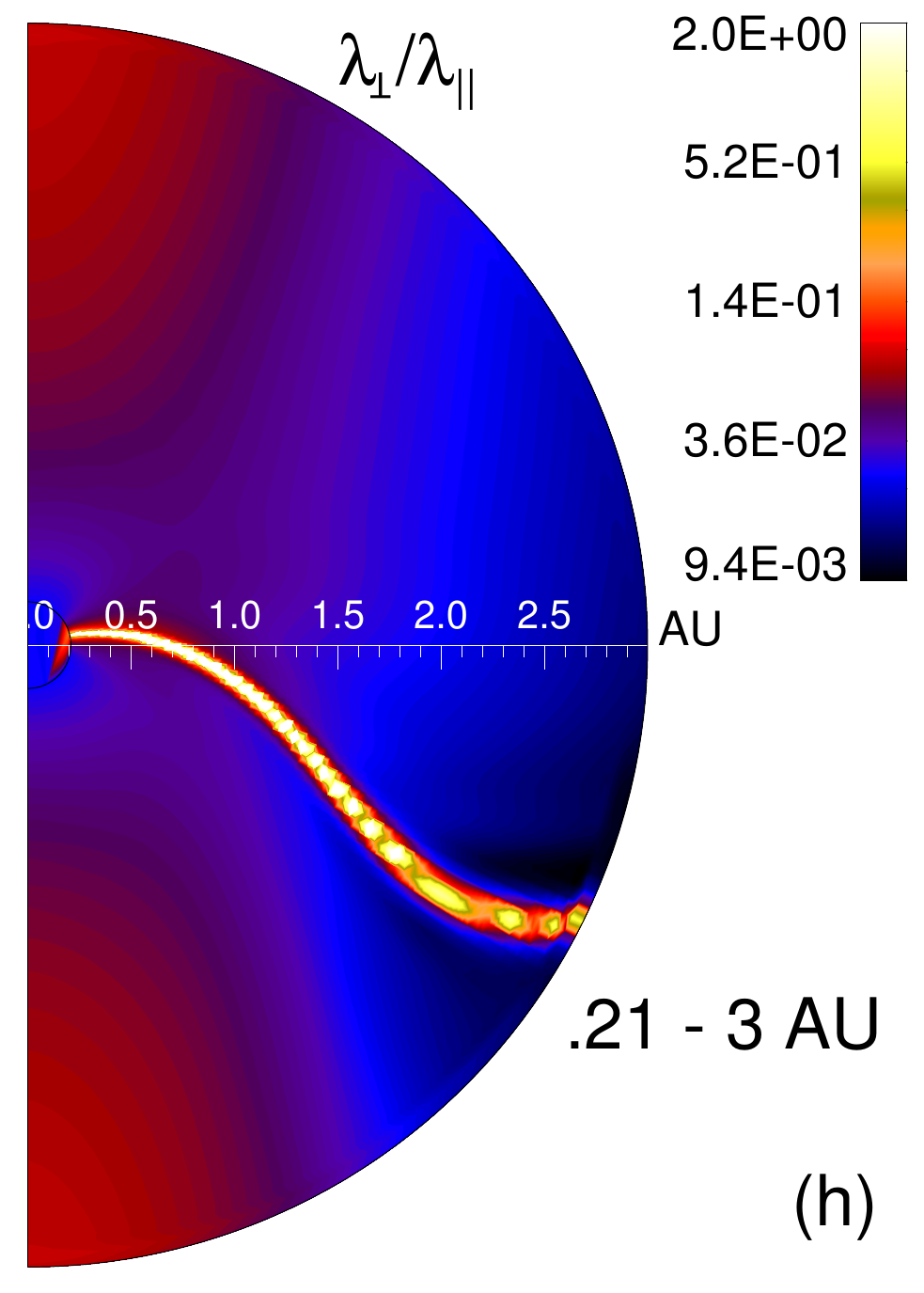}
\caption{Contour plots of mfps in the meridional plane with azimuthal angle of 26\degree, with a solar dipole tilted 30\degree with respect to the solar rotation axis. The inner and intermediate regions ($2 - 45~R_\odot$) and the outer region ($0.21 - 3$ AU, or $45 - 645~R_\odot$) are shown separately. Proton rigidity is 445 MV (100 MeV kinetic energy) and $p=2$.}
\label{fig:merid_tilt}
\end{figure}

Figure \ref{fig:merid_tilt} shows contour plots for mfps in the meridional plane at azimuthal angle equal to $26\degree$, for a simulation with a source magnetic dipole that is tilted by $30\degree$  with respect to the solar rotation axis. In this case, solar rotation produces an asymmetrical magnetic field structure, which has a striking effect on the diffusion parameters, with the displacement of the current sheet from the ecliptic plane modifying their distribution at low latitudes. Note that the blob-like structures in Figures~\ref{fig:merid_tilt}f and \ref{fig:merid_tilt}h arise due to grid points coinciding with the HCS. The rapid decrease in the magnitude of the IMF near the HCS leads to the formation of the blob contours around grid points where $B$ vanishes. This effect is not seen in Figure~\ref{fig:merid_untilt} for the untilted dipole case, where the HCS lies at $0\degree$ heliolatitude, where no grid points are present, by construction.

As noted previously in Section 4.2, observations indicate that the ratio $\lambda_\perp/\lambda_\parallel$ may  approach, and even exceed unity. In our simulation, this happens in the HCS. The basic features described above for the untilted dipole are still present in this case, but are now organized with respect to the tilted HCS. During periods when solar activity levels are high, the warped current sheet is spread out across a larger portion of the heliosphere (Figure \ref{fig:merid_tilt}) compared with the low activity case (untilted dipole, Figure \ref{fig:merid_untilt}), and the HCS is thus more likely to influence CRs. 


\section{Conclusions and Discussion}

We have presented a detailed analysis of the diffusion coefficients for cosmic ray transport in the inner heliosphere. The purpose is to use a well-tested, fully 3-D global simulation of the solar wind, with turbulence modeling, to obtain the heliospheric distribution of the large-scale heliospheric magnetic field, the energy in the turbulent fluctuations, and the correlation scale of the turbulence. This distribution has been coupled with a quasi-linear theory for parallel diffusion, and the recent random ballistic decorrelation interpretation of the non-linear guiding center theory for perpendicular diffusion. The present work extends previous studies on the heliospheric diffusion of cosmic rays by \cite{bieber1995diffusion}, \cite{zank1998radial},and \cite{pei2010cosmic}, but has a stronger focus on the inner heliosphere, with the inner boundary of our simulations at 1 $R_\odot$. Recent complementary work \citep{guo2016ApJ} carries out similar computations of diffusion coefficients for the outer heliosphere.

We find that at the heliospheric current sheet $\lambda_\perp$ can be greater than $\lambda_\parallel$, but usually $\lambda_\parallel$ is 1-2 orders of magnitude larger through most of the inner heliosphere. Very close to the sun ($2~R_\odot$), the strong IMF leads to a large value of $\lambda_\parallel$ ($\sim 0.5$ AU), which initially decreases for several solar radii before increasing with radial distance at low to intermediate latitudes, and becomes nearly constant at the polar regions. $\lambda_\perp$ increases with heliocentric distance throughout the inner heliosphere, and is larger in the polar regions compared to low latitudes. $\lambda_{rr}$ is dominated by $\lambda_\parallel$ through most of the inner heliosphere. However, $\lambda_\perp$ does affect $\lambda_{rr}$ in parts of the near-ecliptic region. Our estimations of $\lambda_\parallel$ near the ecliptic plane at 1 AU show good agreement with the Palmer consensus range of $0.08 - 0.3$ AU. 

At heliocentric distances below 1 AU, we find that the parallel mfp varies with rigidity as $P^{0.33}$ for all rigidities considered here ($10 - 10^4$ MV). Above 1 AU, highly energetic particles begin to resonate with turbulent fluctuations in the energy containing scales, and the rigidity dependence of $\lambda_\parallel$ steepens. The perpendicular mfp is weakly dependent on rigidity. Our results on the rigidity dependence of mfps are consistent with observations.

The mfps are found to be weakly dependent on the type of power spectrum used to represent the large scale fluctuations. This suggests that any attempts to use spacecraft observations of mfps to infer constraints on the ultrascale would be challenging. The effects of solar activity (via a tilted solar dipole and variations of turbulence levels) are also studied, with increased activity leading to stronger perpendicular diffusion and weaker parallel diffusion.

 The model we have adopted for turbulence transport has been thoroughly studied and tested \citep{breech2008turbulence}. More elaborate models, with more transport equations (and more free parameters) are available \citep{Zank2012ApJ745}. In particular, these models include extensions such as dynamically variable residual energy, separate transport equations for slab and 2-D fluctuations, and as many as three distinct dynamically evolving correlation lengths \citep{Oughton2011JGRA116,Zank2017ApJ835}. For the present we forgo the associated additional complication and rely on the present model's ability to account very well for a variety of observations \citep{usmanov2011solar,usmanov2012three,usmanov2014three}. 

We also remark that the turbulent fluctuations we follow dynamically are the quasi-two dimensional fluctuations that we assume are energetically dominant. A variety of studies \citep{matthaeus1990JGR,zank1993nearly,bieber1994proton,bieber1996dominant} are consistent with dominance by quasi-2D turbulence in solar wind turbulence. In the present approach we assumed that the quasi-slab component of the fluctuations, which represent perhaps 20\% of the total fluctuation energy, are a constant fraction of the turbulence energy. Useful extensions have been presented by \cite{Oughton2011JGRA116,Zank2017ApJ835} that adopt somewhat different approaches with the common goal of independently transporting both 2-D and slab-like fluctuations. As noted above, these models find that the radial evolution of 2-D and slab fluctuation energies is not too dissimilar in the inner heliosphere, and therefore our decomposition of the total turbulence energy into slab and 2-D components using a constant ratio appears reasonable. These models also show that in the outer heliosphere (above 3-4 AU), the energy in the slab fluctuations increases with heliocentric distance due to driving by pickup ions, while the 2-D fluctuation energy continues to decrease. As such, studies of CR diffusion in the outer heliosphere would undoubtedly benefit from using a two-component turbulence transport model.

 Such models have been implemented \citep{wiengarten2016ApJ833,Shiota2017ApJ837}, with many differences relative to the present model. For example, the \cite{Shiota2017ApJ837} model has a more elaborate transport formalism, as described above, but neglects the impact of turbulence on the background flow and relies on ad-hoc shear terms instead of fully coupling to the large-scale solar wind solutions. In contrast, we employ a dynamic eddy-viscosity model \citep{usmanov2014three} to achieve this coupling. Clearly no model at present is a complete treatment, and there are advantages and trade-offs in various approaches. We hope to advance our own model with additional refinements in the near future.
 
We anticipate that 3-D calculations of the CR diffusion coefficients in the way we have demonstrated here, employing large scale solar wind solutions with turbulence transport and turbulence modeling, will become increasingly important for realistic energetic particle transport calculations in the future. We also note that related types of diffusion coefficients, such as drag or self-diffusion, may be similarly estimated using adaptations of the above approach, as described briefly in the Appendix. Studies of phenomena such as shock-ensembles and super-events \citep{Mueller-Mellin1986,Kunow1991book}, where several shocks merge to influence energetic particle transport at widely separated locations, would benefit enormously from such 3-D studies in model heliospheres. Our findings of domains where $\lambda_\perp/\lambda_\parallel \geq 1$ may be used to further study the effects of significant perpendicular diffusion, which has been seen to reduce the SEP flux and make it more uniform \citep{zhang2009ApJ}. Additional development at the MHD level will be needed to utilize this kind of tool for explaining observed SEP events associated with transient phenomena such as flares, CMEs and interplanetary shocks \citep{ruffolo2006ApJ,droge2016ApJ,agueda2016ApJ}. In the present paper we have not undertaken specific calculations  employing the diffusion coefficients we obtained using a global model; this is deferred to future work. We anticipate that this approach will be useful in understanding Solar Probe Plus observations of energetic particles near the Sun.
 
As we have now demonstrated that such an approach can provide detailed three dimensional information concerning both MHD transport and particle mean free paths, it becomes clear that what will be needed are improved methods for driving this kind of model with more sophisticated and detailed solar observations. Meanwhile, we are continuing to improve our MHD modeling by building a coronal module that includes a full turbulence transport model, and by further developing the eddy viscosity approach \citep{usmanov2014three}. Future work could also investigate the influence of drifts on CR modulation. To facilitate use of the present data from this model for particle transport calculations of relevance to the current generation energetic particle and Space Weather studies, we are uploading as Supplementary Material the 3-D grids of the diffusion coefficients that were described here.


\section{Appendix}

Here we present an estimation of a general turbulent diffusion coefficient that is based on Taylor's formulation of the problem \citep{taylor1921ProcLonMathSoc}. The diffusion coefficient for the passive transport of any quantity in a turbulent neutral fluid may be approximated by \citep{Choudhuri1998book} 
\begin{equation} \label{eq:taylor_diff}
D_T \approx \frac{1}{3} \langle v^2 \rangle \tau_{\text{cor}},
\end{equation}
where  $\langle v^2 \rangle$ is the mean square turbulent velocity and $\tau_{\text{cor}}$ is the correlation time of the turbulence. By assuming $\langle v^2 \rangle \sim Z^2$, and defining the turbulence correlation length $\lambda \sim  Z \tau_{\text{cor}}$, we rewrite the above equation as

\begin{equation} \label{eq:drag}
D_T \propto Z \lambda.
\end{equation}
Note that any standard diffusion coefficient, drag coefficient, eddy viscosity, or other similar quantity can be expressed in a form similar to Equation~\eqref{eq:drag}, i.e., as a product of a characteristic velocity and a length scale (see, for example, \cite{Tennekes1972book}).

\begin{figure}
\includegraphics[scale=.445]{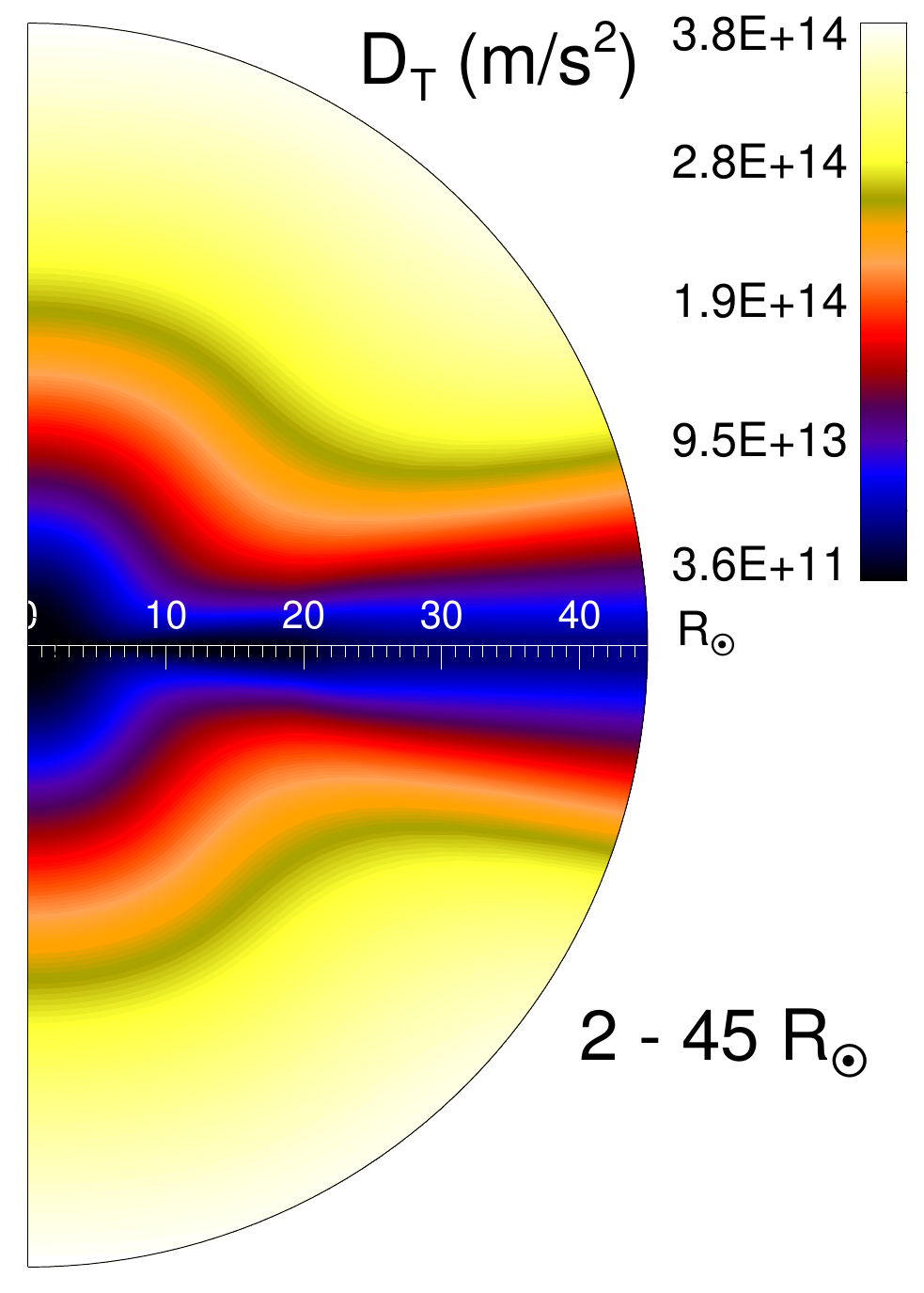}
\includegraphics[scale=.445]{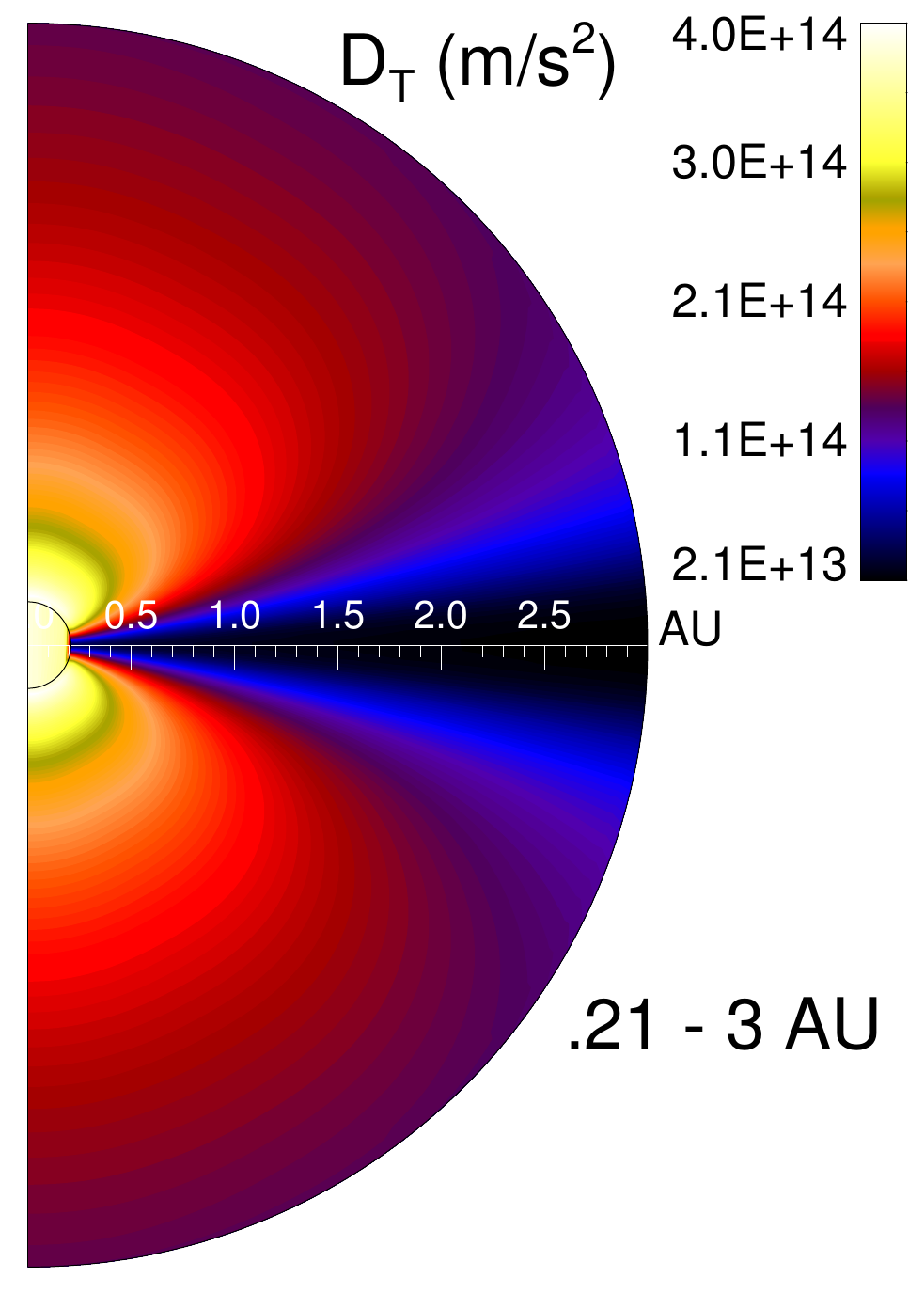}
\caption{Turbulent drag coefficient computed from a simulation with a solar dipole that is untilted with respect to the solar rotation axis. The inner and intermediate regions ($2 - 45~R_\odot$) and the outer region ($0.21 - 3$ AU, or $45 - 645~R_\odot$) are shown separately.}
\label{fig:merid_drag}
\end{figure}

In Figure \ref{fig:merid_drag} we show contour plots for $D_T$ in the meridional plane, computed from a simulation with a solar dipole that is untilted with respect to the solar rotation axis. We may interpret $D_T$ as a turbulent drag coefficient, which is of relevance to the propagation of CMEs in the solar wind. At high heliolatitudes, the drag coefficient increases from the solar surface to 0.5 AU, and then gradually decreases. Notably, at heliocentric distances smaller than 0.5 AU, $D_T$ increases by an order of magnitude in moving from the ecliptic to polar regions. This implies that a CME would be ``channelled" to lower latitudes as it propagates through the inner heliosphere. Applications involving these more general approximations to diffusion processes may also be enabled by the approach described in the present paper. 

\acknowledgments{This research is partially supported by NASA grant NNX14AI63G 
(Heliophysics Grand Challenges Research), NASA LWS grants NNX15AB88G and NNX13AR42G,
and the Solar Probe Plus mission through the ISOIS project and SWRI subcontract D99031L, and the Thailand Research Fund (grant RTA5980003). The authors would like to thank the anonymous Referee for their thorough reading of the manuscript and useful suggestions for its improvement.}

\bibliographystyle{apj}


\end{document}